\documentclass[twocolumn,twocolappendix]{aastex631}
\usepackage{amsmath}

\received{}
\revised{}
\accepted{}

\submitjournal{The Astrophysical Journal}
\shorttitle{Shocks and Energy Dissipation in ISM and ICM}
\shortauthors{Cho et al.}

\begin{document}

\title{Effects of Forcing on Shocks and Energy Dissipation in Interstellar and Intracluster Turbulences}

\author{Hyunjin Cho}
\affil{Department of Physics, College of Natural Sciences UNIST, Ulsan 44919, Korea}
\affil{Department of Earth Sciences, Pusan National University, Busan 46241, Korea}
\author[0000-0002-5455-2957]{Dongsu Ryu}
\affil{Department of Physics, College of Natural Sciences UNIST, Ulsan 44919, Korea}
\author[0000-0002-4674-5687]{Heysung Kang}
\affil{Department of Earth Sciences, Pusan National University, Busan 46241, Korea}

\correspondingauthor{Dongsu Ryu}
\email{dsryu@unist.ac.kr}

\begin{abstract}


Observations indicate that turbulence in the interstellar medium (ISM) is supersonic ($M_{\rm turb}\gg1$) and strongly magnetized ($\beta\sim0.01-1$), while in the intracluster medium (ICM) it is subsonic ($M_{\rm turb}\lesssim1$) and weakly magnetized ($\beta\sim100$). Here, $M_{\rm turb}$ is the turbulent Mach number and $\beta$ is the plasma beta. We study the properties of shocks induced in these disparate environments, including the distribution of the shock Mach number, $M_s$, and the dissipation of the turbulent energy at shocks, through numerical simulations using a high-order accurate code based on the WENO scheme. In particular, we investigate the effects of different modes of the forcing that drives turbulence: solenoidal, compressive, and a mixture of the two. In the ISM turbulence, while the density distribution looks different with different forcings, the velocity power spectrum, $P_v$, on small scales exhibits only weak dependence. Hence, the statistics of shocks depend weakly on forcing either. In the ISM models with $M_{\rm turb}\approx10$ and $\beta\sim0.1$, the fraction of the turbulent energy dissipated at shocks is estimated to be $\sim15~\%$, not sensitive to the forcing mode. In contrast, in the ICM turbulence, $P_v$ as well as the density distribution show strong dependence on forcing. The frequency and average Mach number of shocks are greater for compressive forcing than for solenoidal forcing, so is the energy dissipation. The fraction of ensuing shock dissipation is in the range of $\sim10-35~\%$ in the ICM models with $M_{\rm turb}\approx0.5$ and $\beta\sim10^6$. The rest of the turbulent energy should be dissipated through turbulent cascade.

\end{abstract}

\section{Introduction}\label{sec:s1}

Turbulence prevails in astrophysical flows in a variety of environments. In the interstellar medium (ISM), it is observed on a wide range of scales \citep[see, e.g.,][]{elme2004,mckee2007,hennebelle2012}. Molecular clouds, for instance, contain highly supersonic motions of the turbulent Mach number $M_{\rm turb}\gtrsim10$ on scales larger than $\sim$ 0.1 pc \citep[e.g.,][]{larson1981,solomon1987,heyer2004}, and strong magnetic fields of typically order $\sim$ mG, corresponding to the plasma beta, the ratio of the gas thermal to magnetic pressure, $\beta\sim0.01-1$ \citep[e.g.,][]{crutcher2010,crutcher2012}. In addition, gas motions of $M_{\rm turb}\sim 1$  and a few are observed in the warm ionized medium (WIM) and the cold neutral medium (CNM), respectively \citep[e.g.,][]{tufte1999,heiles2003}. The magnetic field strength in the diffuse ISM is estimated to be several $\mu$G \citep[e.g.,][]{haverkorn2015}, and the plasma beta is $\beta\lesssim1$ or smaller in the WIM and CNM.

It is also well established that the intracluster medium (ICM) is in the state of turbulence. According to X-ray observations of the Coma cluster \citep{shue2004,churazov2012} and the Perseus cluster \citep{hitomi2016}, the typical velocity of turbulent motions is in the range of a few to several $\times100$ km s$^{-1}$. Simulations of cosmic structure formation suggest that the ICM turbulence is subsonic with $M_{\rm turb}\lesssim1$ { \citep[e.g.,][]{ryu2003,ryu2008,vazza2017b,roh2019,moha2020,moha2021}.} Observations of Faraday rotation measures and synchrotron emissions indicate the presence of $\sim\mu$G magnetic fields in the ICM \citep[e.g.,][]{clarke2001,carilli2002,govoni2004}, and structure formation simulations have shown that such magnetic fields can be produced from week seed magnetic fields through small-scale, turbulent dynamo \citep[e.g.,][]{ryu2008,vazza2017a,roh2019}. Then, the plasma beta of the ICM would be of order $\beta \sim 10^2$.

The turbulence in the ISM and ICM is now recognized as one of the key ingredients that govern the properties of the systems. For instance, turbulent diffusion facilitates the transport of mass, momentum, energy, and magnetic fields, and plays important roles in shaping up the physical state \citep[e.g.,][]{brandenburg2011}. In addition, turbulence modifies the density distribution, and hence controls the star formation process in the ISM { \citep[e.g.,][for a review]{maclow2004,federrath2012,krum2019}.} In the ICM, turbulent acceleration is the likely mechanism for the production of cosmic-ray (CR) electrons responsible for the diffuse synchrotron emission of radio halos \citep[see,][and references therein]{brunetti2014}.

Shocks arise naturally, heating the gas, in compressible turbulence. It was suggested that the filamentary structures in the ISM, where protostellar cores are preferentially found, originate from intersecting shocks induced in supersonic turbulence { \citep[e.g.,][]{pudritz2013,federrath2016}.} Shocks can also trigger chemical reactions, driving chemical evolution in the ISM \citep[e.g.,][]{jorgensen2004}. In the ICM, shocks control the production and evolution of vorticity \citep[e.g.,][]{porter2015,vazza2017b} and accelerate CR protons and electrons \citep[e.g.,][]{ryu2003,kang2012,ryu2019}. And some of those are manifested as radio relics \citep[see, e.g.,][]{vanWeeren2019}.

Shocks in astrophysical turbulence have been previously studied using isothermal, compressible, magnetohydrodynamic (MHD) simulations. For instance, \citet{porter2015} obtained the probability distribution function (PDF) of the shock Mach number, $M_s$, for shocks induced in the ICM turbulence with $M_{\rm turb}\approx0.5$ and initial $\beta_0=10^6$. They showed that in turbulent flows, the PDF of $M_s$ follows the power-law form. \citet{lehmann2016} analyzed shocks in the ISM turbulence with $M_{\rm turb}\approx9$ and $\beta\lesssim0.1$. They showed that both fast and slow shocks form in molecular clouds, and slow shocks are as frequent as fast shocks. Recently, \citet[][PR2019, hereafter]{park2019} studied shocks in turbulent media with different parameters, ranging $M_{\rm turb}\approx0.5-7$ and $\beta_0=0.1-10$. In particular, they estimated the amount of the turbulent energy dissipated at shocks, $\epsilon_{\rm shock}$, for the first time, and found that fast shocks are responsible for most of the dissipation and $\epsilon_{\rm shock}$ depends on turbulence parameters. The fraction of the turbulent energy dissipated at shocks, that is, the ratio of the energy dissipated at shocks and the injected energy, is estimated to be $\sim10-40~\%$.

In simulations, while turbulence can be driven with the so-called solenoidal forcing $(\mbox{\boldmath$\nabla$}\cdot\delta\mbox{\boldmath$v$}=0)$, or compressive forcing $(\mbox{\boldmath$\nabla$}\times\delta\mbox{\boldmath$v$}=0)$, or even mixtures of the two, the outcomes depend on the forcings. Naturally, the properties of turbulent flows turn out to be different with different forcings. For example, the density distribution exhibits broader PDFs and more intermittent structures with compressive forcing than with solenoidal forcing \citep[e.g.,][]{federrath2008,federrath2009}. The amplification of magnetic fields through small-scale dynamo is less efficient with compressive forcing \citep[e.g.,][]{federrath2011,porter2015,lim2020}. And shocks are on average stronger with compressive forcing, specially if magnetic fields are weak with $\beta\gg1$ \citep[e.g.,][]{porter2015}.

In this paper, we study ``shocks'' in simulated turbulences that are intended to reproduce the ISM and ICM environments, focusing on the ``effects of different forcing modes''.\footnote{The properties of turbulent flows can depend not only on the forcing mode, $\mbox{\boldmath$\nabla$}\cdot\delta\mbox{\boldmath$v$}=0$ or $\mbox{\boldmath$\nabla$}\times\delta\mbox{\boldmath$v$}=0$, but also on the temporal coherency of turbulence driving, that is, whether the driving vector $\delta\mbox{\boldmath$v$}$ changes continuously over a finite period or $\delta\mbox{\boldmath$v$}$ is drawn randomly at each time step. It was shown that, for instance, the correlation between the density and magnetic field strength differs, if the temporal coherency is different \citep[see][]{yoon2016}. The properties of shocks described in this paper could be affected by the temporal coherency of driving too, although its effects may not be as large as those of forcing mode. In this work, the driving of turbulence is temporally uncorrelated (see Section \ref{sec:s2.2} below), and we do not consider temporally correlated drivings.} Previously, for instance, either solenoidal forcing (e.g., PR2019) or a mixture of solenoidal/compressive forcings \citep[e.g.,][]{lehmann2016} were used for studies of shocks in astrophysical turbulence. Here, we extend the work of PR2019 to include specifically the followings: (1) simulations for the turbulence in ISM molecular clouds with $M_{\rm turb}\approx10$ and $\beta_0=0.1$ and also simulations for the ICM turbulence with $M_{\rm turb}\approx0.5$ and $\beta_0=10^6$, and (2) the quantification of shock properties, such as the Mach number distribution of shocks and the energy dissipation at shocks, for turbulences with three different forcing modes, i.e., solenoidal, compressive, and a mixture of the two. Also in this work, a high-order accurate MHD code, based on the finite-difference, weighted essentially non-oscillatory (WENO) scheme \citep{jiang1996,jiang1999}, is used for simulations.

The paper is organized as follows. In Section \ref{sec:s2}, we describe the numerical setups, including the details of forcing modes. The identification of shocks and the analyses of shock properties are also described. In Section \ref{sec:s3}, we present the results, including the Mach number distribution of shocks and the energy dissipation at shocks in simulated turbulences. A summary and discussion follow in Section \ref{sec:s4}. A brief description of the WENO code used in the work is given in Appendix A.

\section{Numerics}\label{sec:s2}

\subsection{MHD Equations}\label{sec:s2.1}

The dynamics of isothermal, compressible, magnetized gas is described by the following MHD equations.
\begin{eqnarray}
{\partial\rho\over\partial t} + \mbox{\boldmath$\nabla$}\cdot(\rho\mbox{\boldmath$v$}) = 0,~~~~~~~~~~~~~~~~~ \label{eq-rho}\\
{\partial\mbox{\boldmath$v$}\over \partial t} + \mbox{\boldmath$v$}\cdot\mbox{\boldmath$\nabla$}\mbox{\boldmath$v$}
+ {1\over\rho}\mbox{\boldmath$\nabla$}P -{1\over\rho}(\mbox{\boldmath$\nabla$}\times\mbox{\boldmath$B$})\times\mbox{\boldmath$B$} = 0, \label{eq-mom}\\
{\partial\mbox{\boldmath$B$}\over\partial t} - \mbox{\boldmath$\nabla$}\times(\mbox{\boldmath$v$}\times\mbox{\boldmath$B$})=0,~~~~~~~~~~~~~ \label{eq-ene}
\end{eqnarray}
where $\rho$ is the gas density, \mbox{\boldmath$v$} and \mbox{\boldmath$B$} are the velocity and magnetic field vectors. The gas pressure is given as $P=\rho c_{s}^2$ with a constant sound speed $c_{s}$. Note that the unit of \mbox{\boldmath$B$} is chosen so that $4\pi$ does not appear in Equation (\ref{eq-mom}).

For simulations of turbulence, we solve the above equations numerically using an MHD code based on the WENO scheme \citep{jiang1996,jiang1999}. The WENO code used in this work has fifth-order spatial and fourth-order temporal accuracies, respectively, while the TVD code used previously in PR2019 is second-order accurate in both space and time. Appendix A briefly describes the WENO code, including a comparison with the TVD code. Viscous and resistive dissipations are not explicitly included. Simulations are performed in a three-dimensional (3D) cubic box of the size $L_0 = 1$ with $256^3$ and $512^3$ uniform Cartesian grid zones. The background medium is initialized with $\rho_0 =1$, $P_0=1$ (i.e., $c_{s}=1$), and $\mbox{\boldmath$v$}_0=0$. The initial magnetic field is placed along the $x$-axis with the uniform strength $B_0$, which is specified by the plasma beta, $\beta_0 = P_{0}/P_{\rm B,0}=2c^2_s\rho_0/B_0^2$. We adopt $\beta_0 =0.1$ for ISM turbulence and $\beta_0 =10^6$ for ICM turbulence.

\begin{deluxetable*}{lcccccccccccccc}
\tablecaption{Model Parameters of Simulations and Statistics of Turbulence$^a$\label{tab:t1}}
\tabletypesize{\small}
\tablecolumns{13}
\tablenum{1}
\tablewidth{0pt}
\tablehead{
\colhead{Model} &
\colhead{$M_{\rm turb}$} &
\colhead{$\beta_0$$^b$} &
\colhead{Forcing} &
\colhead{${t_{\rm end}}/{t_{\rm cross}}$$^c$} &
\colhead{$\epsilon_{\rm inj}$$^d$} &
\colhead{$\beta_{\rm sat}$$^b$} &
\colhead{$({N_{\rm fa}}/{n_{g}^2})$$^e$} &
\colhead{$({N_{\rm sl}}/{n_{g}^2})$$^e$} &
\colhead{$\epsilon_{\rm fa}$$^d$} &
\colhead{$\epsilon_{\rm sl}$$^d$} &
\colhead{${\epsilon_{\rm shock}}/{\epsilon_{\rm inj}}$}
}
\startdata
ISM-Sol-N256     & 10  & 0.1      & Solenoidal     & 5     & 1650     & 0.0407                        & 4.69 & 0.640   & 241    & 2.63    & 0.148 \\
ISM-Sol-N512     & 10  & 0.1      & Solenoidal     & 2.2   & 1590     & 0.0388                       & 5.35& 1.40    & 207    & 5.62     & 0.134 \\
\hline
ISM-Mix-N256    & 10  & 0.1      & Mixed           & 5      & 1550     & 0.0429                       & 4.66 & 0.612  & 234     & 2.70    & 0.153 \\
ISM-Mix-N512    & 10  & 0.1      & Mixed           & 3.5    & 1520     & 0.0419                       & 5.35 & 1.42   & 201     & 5.79    & 0.136 \\
\hline
ISM-Comp-N256   & 10  & 0.1      & Compressive & 5      & 1300    & 0.0564                      & 3.43 & 0.438 & 181     & 2.37   & 0.141 \\
ISM-Comp-N512   & 10  & 0.1      & Compressive & 3.5   & 1200     & 0.0550                      & 4.93 & 1.01   & 185     & 4.58   & 0.158 \\
\hline
ICM-Sol-N256     & 0.5 & $10^6$ & Solenoidal     & 30    & 0.140    & 5.31 $\times 10^{1}$  & 6.03 & 0.0     & 0.0139 & 0.0    & 0.100 \\
ICM-Sol-N512     & 0.5 & $10^6$ & Solenoidal     & 30    & 0.140     & 4.71 $\times 10^{1}$ & 6.43 & 0.0     & 0.0150 & 0.0    & 0.107 \\
\hline
ICM-Mix-N256    & 0.5 & $10^6$ & Mixed           & 30    & 0.150    & 5.79 $\times 10^{1}$ & 8.18 & 0.0      & 0.0281 & 0.0    & 0.188 \\
ICM-Mix-N512    & 0.5 & $10^6$ & Mixed           & 30    & 0.146    & 5.12 $\times 10^{1}$ & 8.38  & 0.0      & 0.0281 & 0.0    & 0.193 \\
\hline
ICM-Comp-N256 & 0.5 & $10^6$ & Compressive & 30    & 0.340     & 8.76 $\times 10^{2}$ & 10.7 & 0.0     & 0.119  & 0.0     & 0.349 \\
ICM-Comp-N512 & 0.5 & $10^6$ & Compressive & 30    & 0.340     & 5.59 $\times 10^{2}$ & 11.7 & 0.0     & 0.113 & 0.0     & 0.331 \\
\enddata
\tablenotetext{a}{The statistics of turbulence, $\beta_{\rm sat}$, $N_{\rm fa}$, $N_{\rm sl}$, $\epsilon_{\rm fa}$, $\epsilon_{\rm sl}$, and $\epsilon_{\rm shock}$, are the mean values at saturation, which are calculated over $2t_{\rm cross}\leq t\leq t_{\rm end}$ for the ISM models and over $15t_{\rm cross}\leq t\leq t_{\rm end}$ for the ICM models. Here, the subscripts ``fa'' and ``sl'' stand for fast and slow shocks, respectively, and $\epsilon_{\rm shock}=\epsilon_{\rm fa}+\epsilon_{\rm sl}$.}
\tablenotetext{b}{The initial plasma beta, $\beta_0$, and the plasma beta at saturation, $\beta_{\rm sat}$.}
\tablenotetext{c}{The end time of simulations in units of the crossing time, $t_{\rm cross}=L_{\rm inj}/v_{\rm rms}$ (see the main text).}
\tablenotetext{d}{The energy injection rate and the energy dissipation rate at shocks in computational units of $\rho_0=1$, $c_s=1$, and $L_0=1$.}
\tablenotetext{e}{The numbers of shock zones normalized to $n_{g}^2$.}
\end{deluxetable*}

\subsection{Turbulence Forcing}\label{sec:s2.2}

Turbulence is driven by adding a small velocity perturbation $\delta\mbox{\boldmath$v$}$ at each grid zone at each time step; $\delta\mbox{\boldmath$v$}$ is drawn from a Gaussian random distribution with a Fourier power spectrum, 
$|\delta\mbox{\boldmath$v$}_k|^2\propto k^6 {\rm exp}(-8k/k_{\rm exp})$, where $k_{\rm exp}=2k_0$ with $k_0=2 \pi /L_0$  \citep{stone1998,maclow1999}. The injection scale is regarded as the peak of $|\delta\mbox{\boldmath$v$}_k|^2k^2$, $k_{\rm inj}=k_{\rm exp}$, i.e., $L_{\rm inj}=L_0/2$. The perturbations have random phases, hence the driving is temporally uncorrelated. The amplitude of $\delta\mbox{\boldmath$v$}$ is adjusted, so that turbulence saturates with the root-mean-square (rms) velocity of flow motions, $v_{\rm rms} = \langle v^2 \rangle^{1/2} \approx M_{\rm turb} c_{s}$. We aim to obtain $M_{\rm turb}=10$ for supersonic ISM turbulence in molecular clouds, and $M_{\rm turb}=0.5$ for subsonic ICM turbulence.

Forcing with $\delta\mbox{\boldmath$v$}$ generally leads to a combination of solenoidal and compressive components. By separating the two components in Fourier space, we construct three types of forcings: (1) fully solenoidal forcing with $\mbox{\boldmath$\nabla$}\cdot\delta\mbox{\boldmath$v$}=0$, (2) fully compressive forcing with $\mbox{\boldmath$\nabla$}\times\delta\mbox{\boldmath$v$}=0$, and (3) a mixture of the two. It was argued that the driving of turbulence in ISM molecular clouds could be specified by a mixture of solenoidal and compressive modes with the 2:1 ratio \citep{federrath2010}. Hence, we consider the case of 2:1 mixed forcing.

Table \ref{tab:t1} summarizes the basic parameters of turbulence models considered in this paper. The nomenclature for the models have three elements, as listed in the first column, which are self-explanatory: (1) the first element indicates either the ``ISM'' or ``ICM'' turbulence, (2) the second element denotes the forcing mode, and (3) the third element shows the grid resolution, N$n_{g}$, where $n_{g}$ is the number of grid zones in one side of the simulation box. 
The high-resolution models {specified with N512 have $512^3$ grid zones, while the low-resolution models specified with N256 have $256^3$ grid zones. The models are defined by the two parameters, $M_{\rm turb}$ (column 2) and $\beta_0$ (column 3), and the forcing mode (column 4). Simulations for the ISM models were run up to $t_{\rm end}=5~t_{\rm cross}$ for N256 models and $t_{\rm end}=2.2-3.5~t_{\rm cross}$ for N512 models, whereas those for the ICM models run up to $t_{\rm end}=30~t_{\rm cross}$ regardless of the resolution. Here, $t_{\rm cross}=L_{\rm inj}/v_{\rm rms}$ is the crossing time. In the weakly magnetized ICM turbulence, the magnetic field amplification via small-scale dynamo is much slower in units of $t_{\rm cross}$, and hence the time scale to reach saturation is much longer than in the ISM turbulence (see Figure \ref{fig:fig1} and the discussion in the next section). The values of $t_{\rm end}/t_{\rm cross}$ are listed in the fifth column.

\subsection{MHD Shocks}\label{sec:s2.3}

In MHDs, there are two kinds of shocks, i.e., fast-mode and slow-mode shocks. The shock identification scheme and the formulas used for the calculations of the shock Mach number and the energy dissipation at shocks are basically same as those in PR2019, except that we here identify shocks with $M_s\geq1.05$, rather than $M_s\geq1.06$, taking the advantage of the high-order accuracy of the WENO code. 

``Shock zones'' are found by a dimension-by-dimension identification scheme, as follows. (1) Along each coordinate direction, grid zones are marked as ``shocked'', if $\mbox{\boldmath$\nabla$}\cdot\mbox{\boldmath$v$}<0$ and $\max(\rho_{i+1}/\rho_{i-1},\rho_{i-1}/\rho_{i+1}) \geq 1.02^2$ around the zone $i$. (2) Considering that shocks spread typically over two to three grid zones, the zone with minimum $\mbox{\boldmath$\nabla$}\cdot\mbox{\boldmath$v$}$ among attached shocked zones is tagged as a ``shock zone''. (3) Around the shock zone, the preshock or postshock zones are chosen, depending on the density jump. (4) The Mach number of shock zones, $M_s$, is calculated using the preshock and postshock quantities (see below). If a zone is identified as shock more than once along coordinate directions, $M_s$ is determined as $M_s=\max(M_{s,x},M_{s,y},M_{s,z})$.

The formula for $M_{s}$ can be derived using the shock jump condition for isothermal MHD flows from Equations (\ref{eq-rho})-(\ref{eq-ene}). Following PR2019, we use
\begin{equation}
M_{s}^2 = \chi+\frac{\chi}{\chi-1}\frac{B_2^2-B_1^2}{2 c_{s}^2 \rho_1}, 
\label{eq-ms}
\end{equation}
where $\chi={\rho_2}/{\rho_1}$. Hereafter, the subscripts `1' and `2' denote the preshock and postshock states, respectively. The second term in the right hand side represents the magnetic field contribution to $M_{s}$. We point out that both $B_2^2-B_1^2$ in the numerator and $\chi-1$ in the denominator goes to zero as $M_s$ approaches to unity in weak shocks. In the ISM turbulence with $M_{\rm turb}\approx10$, most of shocks are strong with $M_s\gg1$, and hence this does not pose a problem. On the other hand, in the ICM turbulence with $M_{\rm turb}\approx0.5$, the shock population is dominated by weak shocks, and uncertainties may be introduced in the PDF of $M_s$ and also the estimate of the energy dissipation at shocks. Equation (\ref{eq-ms}) can be rewritten as
\begin{equation}
M_{s}^2 = \chi+\frac{\chi B_{\perp1}\left(B_{\perp2}+B_{\perp1}\right)}{2c_s^2\rho_1}\frac{\eta}{\zeta},
\label{eq-ms2}
\end{equation}
with $\eta=B_{\perp2}/B_{\perp1}-1$ and $\zeta=\rho_2/\rho_1-1$. Hereafter, the subscripts $\perp$ and $\parallel$ denote the magnetic field components perpendicular and parallel to the shock normal, respectively. It is the ratio, $\eta/\zeta$, that may not be accurately reproduced from the numerical solutions for weak shocks. We note that $\eta/\zeta$ is given as $M_s^2/(M_s^2-\chi v_{{A}\parallel1}^2/c_s^2)$, where $v_{{A}\parallel1}={B_{\parallel}}/{\sqrt{\rho_1}}$. Hence, in the ICM turbulence with $\beta\gg1$, we use Equation (\ref{eq-ms2}), assuming $\eta/\zeta\approx1$, for the calculation of $M_s$, if the magnetic field around shock zones is weak, specifically, if $v_{{A}\parallel1}^2/c_s^2\leq0.1$; if $v_{{A}\parallel1}^2/c_s^2>0.1$, we use Equation (\ref{eq-ms}).

Identified shock zones are classified into either fast or slow shocks, according to the criterion of $B_{\perp2} > B_{\perp1}$ or $B_{\perp2} < B_{\perp1}$. In addition, fast shocks should satisfy $M_s^2 > \chi v_{{A}\parallel1}^2/c_s^2$, while slow shocks satisfy $M_s^2 < v_{{A}\parallel1}^2/c_s^2$. In our simulated turbulence, about 95\% of identified shock zones satisfy these conditions for either fast or slow shocks. The rest have $v_{{A}\parallel1}^2/c_s^2< M_s^2 < \chi v_{{A}\parallel1}^2/c_s^2$, and may be classified as ``intermediate shocks'', which are known to be nonphysical \citep[e.g.,][]{landau1960}. The presence of those intermediate shocks could be partly due to our dimension-by-dimension approach for shock identification and also partly due to possible numerical errors in capturing shocks in simulations. In the next section, we present the results excluding intermediate shocks. With the fraction of about 5\% or so, the exclusion should not affect the main conclusions of this work.

Although the results are presented with the sonic Mach number of shocks, $M_s$, in the next section, the fast or slow Mach numbers can be calculated as $M_{\rm fa}=M_s c_s/c_{\rm fa}$ or $M_{\rm sl}=M_s c_s/c_{\rm sl}$, using the fast and slow wave speeds in the preshock region,
\begin{eqnarray}
c_{\rm fa,sl}^2&=&\frac{1}{2}\left(c_s^2+v_{{A}\parallel1}^2+v_{{A}\perp1}^2\right) \nonumber  \\  
&&\pm\frac{1}{2}\sqrt{\left(c_s^2+v_{{A}\parallel1}^2+v_{{A}\perp1}^2 \right)^2-4v_{{A}\parallel1}^2c_s^2}, \label{eq-cfcs}
\end{eqnarray}
where $v_{{A}\perp1}={B_{\perp1}}/{\sqrt{\rho_1}}$. Hereafter, the subscripts ``fa'' and ``sl'' denote fast and slow shocks, respectively. Considering that weak shocks could be confused with waves, we examine only fast and slow shocks with $M_{\rm fa}\geq1.05$ and $M_{\rm sl}\geq1.05$, respectively, and also with $M_s\geq1.05$. In addition, another constraint $c_{\rm sl}/c_s\geq0.3$ is imposed for slow shocks, since slow shocks with very small $c_{\rm sl}$ are not clearly distinguished from fluctuations.

{ 
We also calculate the energy dissipation at shocks, as in PR2019. With the ``heat energy'' or the ``effective internal energy'', $P\ln P$, the equation for the ``effective total energy'' can be written as
\begin{eqnarray} 
&&\frac{\partial}{\partial{t}}\left(\frac{1}{2}\rho v^2+P\ln P+\frac{1}{2}B^2\right) \nonumber \\
&&+~\mbox{\boldmath$\nabla$}\cdot\left[\left(\frac{1}{2}\rho v^2+P\ln P+P\right)\mbox{\boldmath$v$} + \left(\mbox{\boldmath$B$}\times\mbox{\boldmath$v$}\right)\times\mbox{\boldmath$B$}\right] \nonumber \\
&& = 0.  \label{eff_totE}
\end{eqnarray}
 \citep{mouschovias1974}. Then, the jump of the total energy flux in the shock-rest frame is given as
\begin{eqnarray}
&&\left[\left(\frac{1}{2}\rho v^2+P\ln P+P+B^2\right)v_{\parallel}-B_{\parallel}\left(B_{\parallel}v_{\parallel}+B_{\perp}v_{\perp}\right)\right]^1_2 \nonumber\\
&&\equiv Q,
\end{eqnarray}
where $\left[f\right]^1_2=f_1-f_2$ denotes the difference between the preshock and postshock quantities. Here, $Q$ is the energy dissipation rate per unit area at shock surface, which can be expressed as
\begin{eqnarray}
&&\frac{Q}{\rho_1 M_{s} c_{s}^3} = \nonumber \\
&&\frac{1}{2}M_{s}^2\left[1-\frac{1}{\chi^2}+\frac{v_{{\rm A}\perp1}^2\left(\chi-1\right)\left\{v_{{\rm A}\parallel}^2\left(\chi+1\right)-2M_{s}^2c_{s}^2\right\}}{\left(v_{{\rm A}\parallel}^2\chi - M_{s}^2 c_{s}^2\right)^2}\right] \nonumber \\
&&-\ln \chi.
\label{eq-Q} 
\end{eqnarray}
}

The dissipation rate of the turbulent energy at all shocks inside the entire simulation box is estimated as
\begin{equation}
\epsilon_{\text{fa(sl)}}= \sum_{\text{fast(slow) shocks},j}^{} Q_j (\Delta x)^2,
\label{eq-efasl}
\end{equation}
for either the fast or slow shock populations. Here, $\Delta x=L_0/n_g$ is the size of grid zones, and the summation goes over all the identified shock zones.

This rate is compared to the injection rate of the energy deposited in the simulation box with the forcing described above:
\begin{equation} 
\epsilon_{\rm inj} = \frac{1}{\Delta t}\int \frac{1}{2}\rho\left[\left(\mbox{\boldmath$v$}+\delta\mbox{\boldmath$v$}\right)^2-\mbox{\boldmath$v$}^2\right] dV, \label{eq-inj}
\end{equation}
where $\Delta t$ is the simulation time step. As mentioned above, the amplitude of $\delta\mbox{\boldmath$v$}$ is the adjustable parameter, and the resulting values of $\epsilon_{\rm inj}$ are given in the sixth column of Table \ref{tab:t1}.

\begin{figure*}
\centerline{\includegraphics[width=0.9\textwidth]{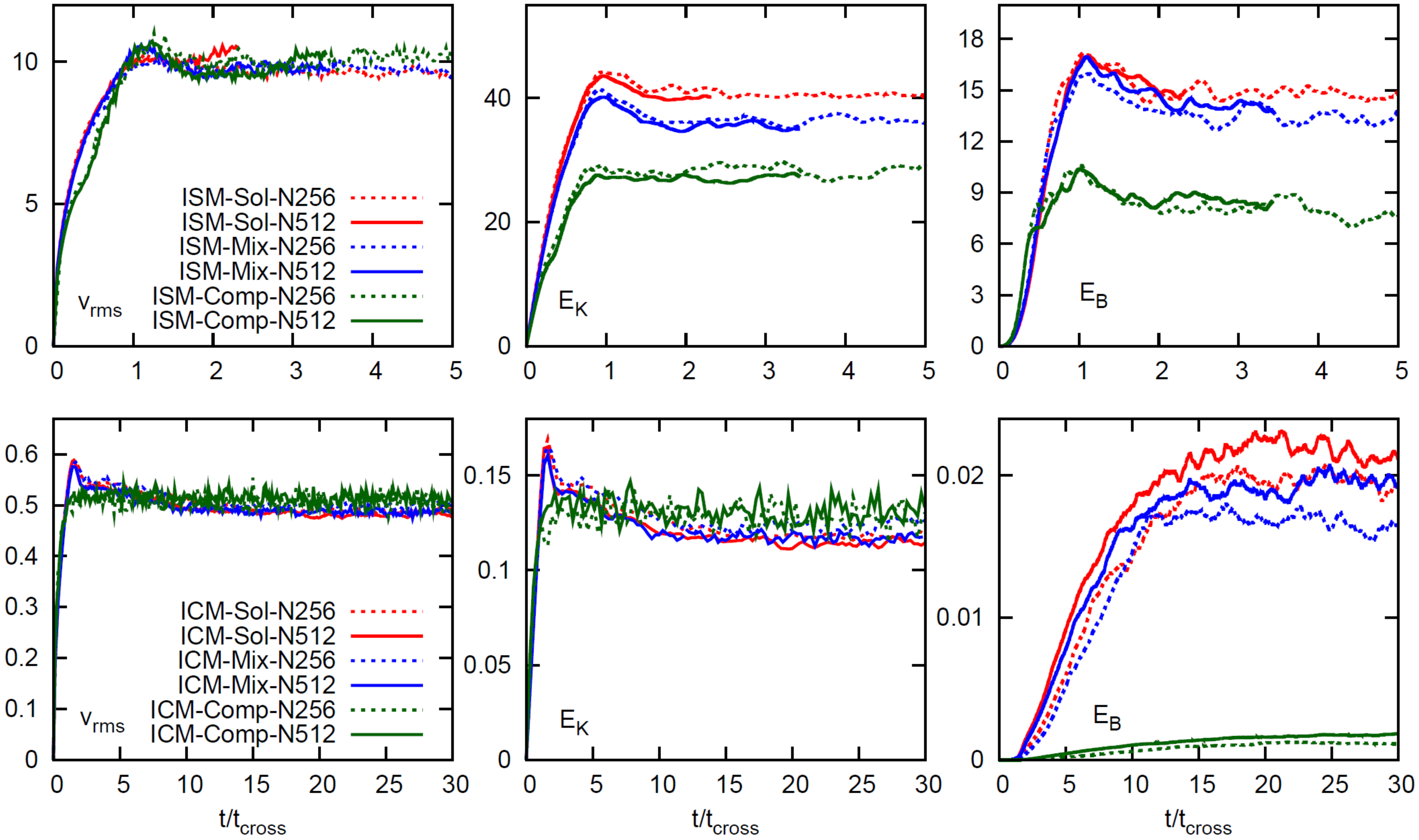}}
\vskip -0.1cm
\caption{Time evolution of the rms velocity, $v_{\rm rms}$ (left), the kinetic energy, $E_K=\int(1/2)\rho v^2 dV$ (middle), and the magnetic energy increase, $E_B=\int(1/2)\delta B^2 dV$ where $\delta\mbox{\boldmath$B$}=\mbox{\boldmath$B$}-\mbox{\boldmath$B_0$}$ (right), for the ISM turbulence models (upper panels) and the ICM turbulence models (lower panels). Each panel contains lines for six models whose parameters are given in Table \ref{tab:t1}. Throughout the paper, the line plots are color-coded, according to the forcing modes: red, blue, and green for the solenoidal, mixed, and compressive forcing models, respectively. The solid lines show the results of the N512 models, while the dotted lines show the results of the N256 models.}
\label{fig:fig1}
\end{figure*}

\begin{figure*}
\centerline{\includegraphics[width=1\textwidth]{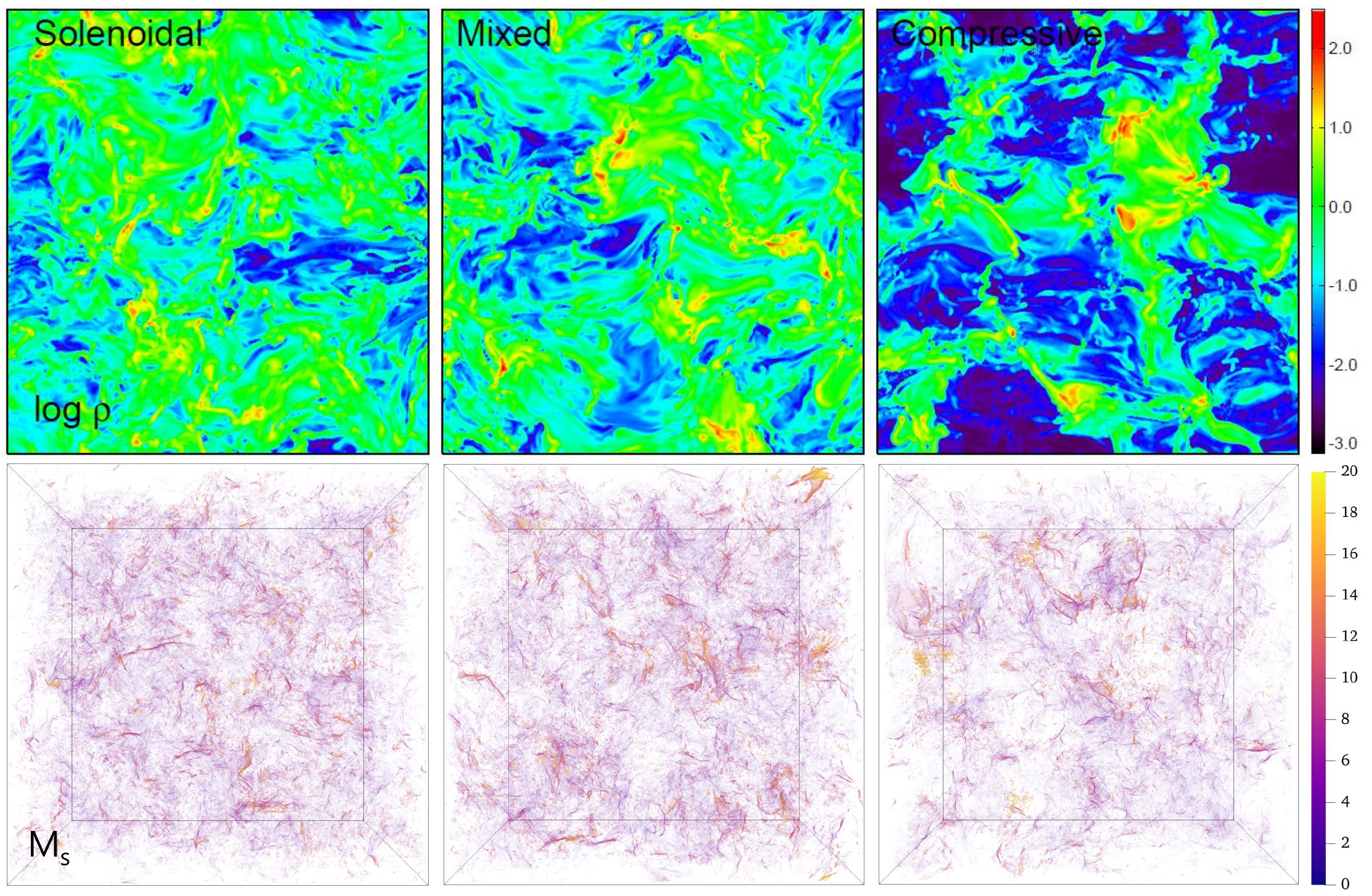}}
\vskip -0.1cm
\caption{2D slice images of $\log \rho/\rho_0$  (upper panels) and 3D distributions of shock zones  color-coded by $M_s$ (lower panels) for the ISM turbulence models. Three high-resolution models, ISM-Sol-N512 (left), ISM-Mix-N512 (middle), and ISM-Comp-N512 (right), at $t=t_{\rm end}$ are shown. See Table \ref{tab:t1} for the model parameters.}
\label{fig:fig2}
\end{figure*}

\section{Results}\label{sec:s3}

\subsection{Evolution and Saturation of Turbulence}\label{sec:s3.1}

We first examine the overall behavior of turbulence in our simulations. Figure \ref{fig:fig1} shows the rms velocity of turbulent flow motions, $v_{\rm rms}$, the volume-integrated kinetic energy, $E_K=\int(1/2)\rho v^2 dV$, and the volume-integrated turbulent magnetic energy, $E_B=\int(1/2)\delta B^2 dV$, as a function of time, for all the models listed in Table \ref{tab:t1}. Here, $\delta\mbox{\boldmath$B$}=\mbox{\boldmath$B$}-\mbox{\boldmath$B_0$}$. Throughout the paper, the turbulence quantities are given in computational units of $\rho_0=1$, $c_s=1$, and $L_0=1$, unless otherwise specified. Regardless of forcings, $v_{\rm rms}\approx 10$ and $0.5$ is attained at saturation in the ISM and ICM turbulence models, respectively.

In the ISM turbulence, $v_{\rm rms}$ as well as $E_K$ and $E_B$ grow until $\sim t_{\rm cross}$, and then after some adjustments, reach saturation by $\sim 2~t_{\rm cross}$. For the models with the same $M_{\rm turb}$, both $E_K$ and $E_B$ are smaller in the compressive forcing models than in the solenoidal forcing models. This is because the density distribution is more intermittent \citep[e.g.,][]{federrath2008,federrath2009} and the small-scale dynamo is less efficient \citep[e.g.,][]{federrath2011,lim2020} with compressive forcing. The averaged values of $\beta$ at saturation ($2~t_{\rm cross}\leq t\leq t_{\rm end}$), $\beta_{\rm sat}$, is listed in the seventh column of Table \ref{tab:t1}. It indicates that the magnetic energy grows more than by a factor of two in the solenoidal forcing models, while the growth factor is somewhat less than two in the compressive forcing models. The mixed forcing models show the behaviors in between.

In the ICM turbulence, reaching saturation takes longer in term of $t_{\rm cross}$, as mentioned in Section \ref{sec:s2.2}. In particular, initially with a very weak seed, the growth of magnetic field needs a number of eddy turn-overs, and hence $E_B$ approaches saturation after $t\sim15~t_{\rm cross}$. Moreover, the saturated $E_B$ is much smaller in the compressive forcing models (green lines) than in the solenoidal forcing models (red lines), since the solenoidal component of flow motions, which is responsible for most of magnetic field amplification, is smaller with compressive forcing, as shown before in \citet{porter2015}. The saturated $E_B$ for the mixed forcing models (blue lines) is a bit smaller than, yet similar to, that for the solenoidal forcing models. As a consequence, $\beta_{\rm sat}$ at saturation ($15~t_{\rm cross}\leq t\leq t_{\rm end}$) is close to $\sim50$ in the solenoidal and mixed forcing models, while it is an order of magnitude larger in the compressive forcing models (see the seventh column of Table \ref{tab:t1}). Considering that $\beta$ in the ICM is estimated to be $\sim50-100$ \citep[e.g.,][]{ryu2008,brunetti2014}, the solenoidal and mixed forcing models would represent more realistic ICM turbulence. Another notable point is that $E_B$ in the N512 models is somewhat larger than $E_B$ in the N256 models. This tells that the amplification of magnetic field by small-scale dynamo is sensitive to the effective Reynolds and Prandtl numbers, which are controlled by numerical resolution in our simulations \citep[see also, e.g.,][]{schober2012,roh2019}. In our ICM-Sol-N512 model, $E_B/E_K$ approaches $\sim20\%$ at saturation, while $E_B/E_K\sim30\%$ was obtained in a $2048^3$ simulation using the TVD code in \citet{porter2015}.

For the analysis of turbulent flows in the following subsections, we examine the mean quantities at saturation, which are calculated over $2~t_{\rm cross}\leq t\leq t_{\rm end}$ for the ISM models and $15~t_{\rm cross}\leq t\leq t_{\rm end}$ for the ICM models.

\subsection{ISM Turbulence}\label{sec:s3.2}

\begin{figure}
\hskip -0.1cm
\centerline{\includegraphics[width=0.5\textwidth]{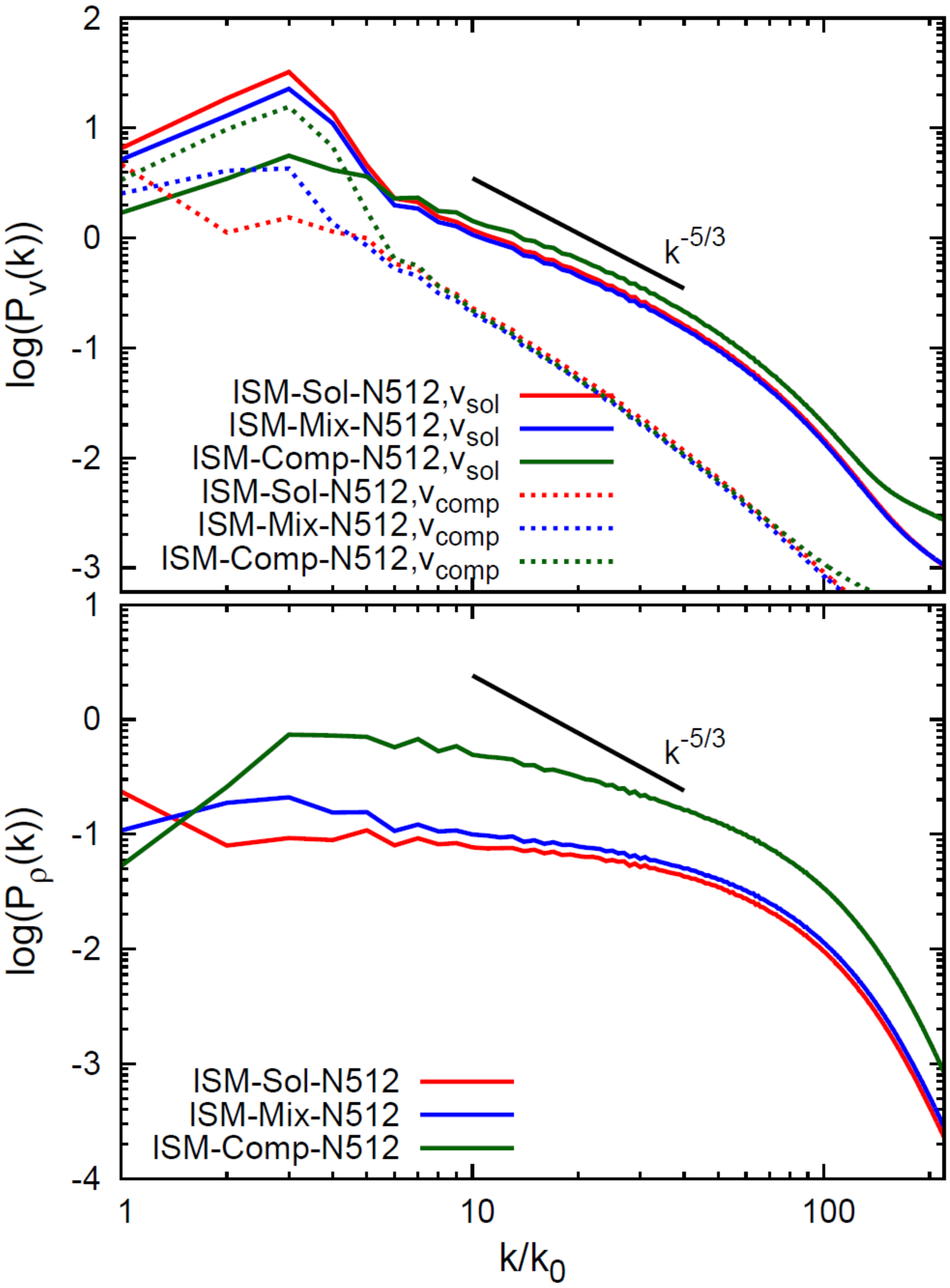}}
\vskip -0.2cm
\caption{{\it Upper panel:} Time-averaged power spectra of the solenoidal component of $\mbox{\boldmath$v$}$, $P_v^{\rm sol}(k)$ (solid lines), and the compressive component of $\mbox{\boldmath$v$}$, $P_v^{\rm comp}(k)$ (dotted lines), for the ISM turbulence models. {\it Lower panel:} Time-averaged power spectra of the density, $P_{\rho}(k)$, for the ISM turbulence models. The power spectra are calculated using the quantities at saturation ($2t_{\rm cross}\leq t\leq t_{\rm end}$). Three high-resolution models are shown. The black lines draw the Kolmogorov spectrum for comparison.}
\label{fig:fig3}
\end{figure}

\begin{figure*}
\centerline{\includegraphics[width=0.9\textwidth]{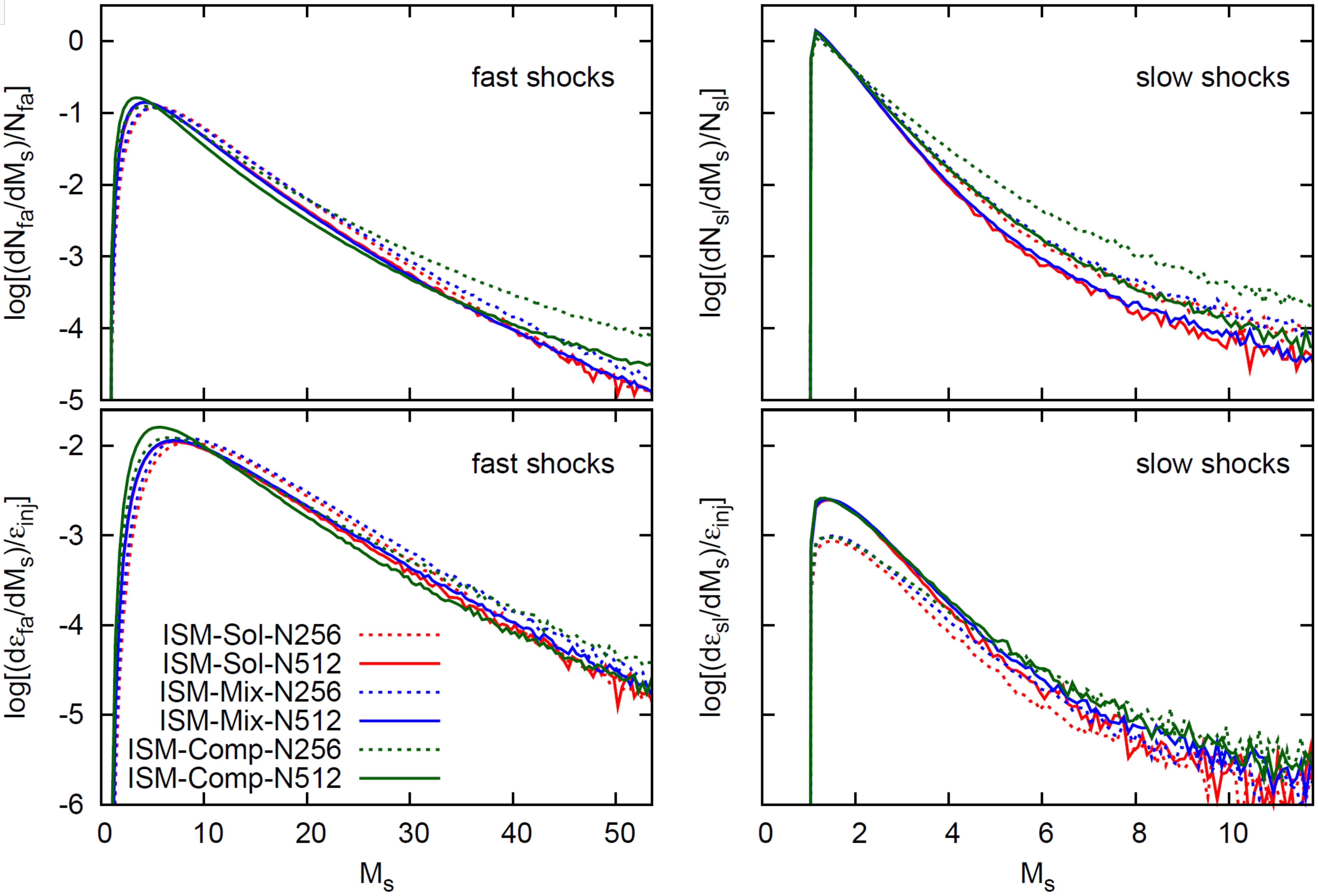}}
\vskip -0.1cm
\caption{Time-averaged PDFs of $M_s$ (upper panels) and time-averaged energy dissipation rates at shocks as a function of $M_s$ (lower panels) for fast shocks (left panels) and slow shocks (right panels) in the ISM turbulence models. The distributions are calculated using the quantities at saturation ($2t_{\rm cross}\leq t\leq t_{\rm end}$). All the six ISM models in Table \ref{tab:t1} are shown.}
\label{fig:fig4}
\end{figure*}

\begin{figure}
\hskip -0.1cm
\centerline{\includegraphics[width=0.5\textwidth]{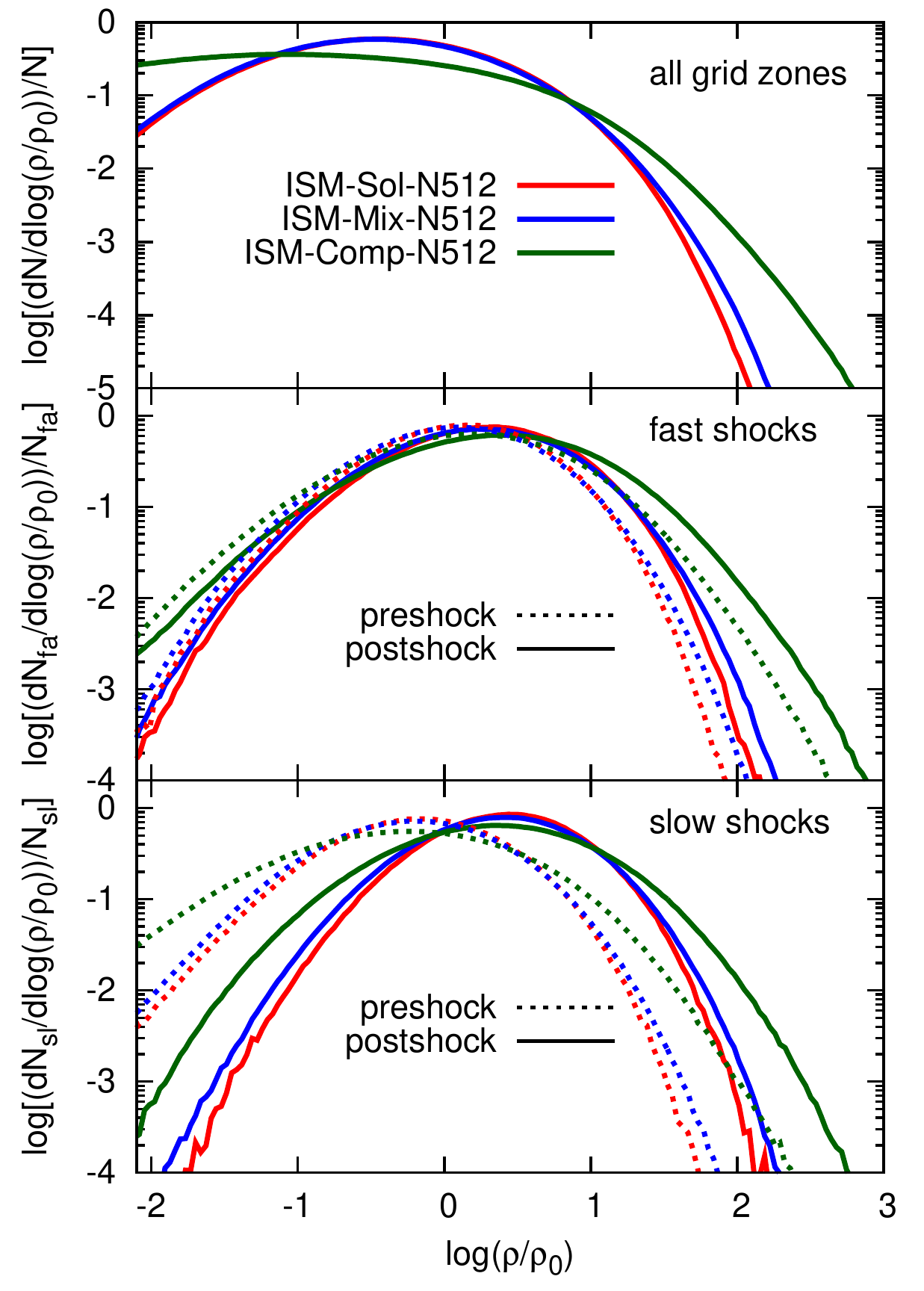}}
\vskip -0.2cm
\caption{Time-averaged PDFs of the gas density in all the computational volume (top panel) and the preshock and postshock gas densities for fast shocks (middle panel) and slow shocks (bottom panel) in the ISM turbulence models. The PDFs are calculated using the quantities at saturation ($2t_{\rm cross}\leq t\leq t_{\rm end}$). Three high-resolution ISM models are shown.}
\label{fig:fig5}
\end{figure}

Figure \ref{fig:fig2} shows the two-dimensional (2D) slice maps of the density (upper panels) and the 3D distributions of shock zones color-coded by $M_s$ (lower panels) in the N512 ISM models with three different forcings at $t=t_{\rm end}$. The density images clearly exhibit the characteristic morphologies with different forcings; the density has a larger contrast and a higher intermittency in the compressive forcing model (right panel) than in the solenoidal forcing model (left panel), which is consistent with previous studies using hydrodynamic simulations \citep[e.g.,][]{federrath2008,federrath2009}. A complex network of shocks appears, regardless of forcings. While shocks are distributed relatively uniformly throughout the entire simulation box in the solenoidal forcing model, the distribution is more concentrated in the compressive forcing model. Again the mixed forcing model (middle panel) shows the behaviors in between. We note that the shock zones include both fast and slow shocks, while strong shocks with high $M_s$ are mostly fast shocks \citep[see the discussion below, and also][and PR2019]{lehmann2016}.

Figure \ref{fig:fig3} shows the power spectra of the flow velocity and density, averaged over the saturated period, for the three N512 ISM models. In the lower panel, $P_{\rho}(k)$ exhibits a clear dependence on forcing. $P_{\rho}(k)$ is several to an order of magnitude larger in the compressive forcing model than in the solenoidal forcing model. $P_{\rho}(k)$ is a bit larger in the mixed forcing model than in the solenoidal forcing model. The difference in $P_{\rho}(k)$ should reflect the visual impression of the density slice images in Figure \ref{fig:fig2}. In addition, $P_{\rho}(k)$ has slopes flatter than the Kolmogorov slope, $-5/3$, in all the models with different forcings, and this is a characteristic property of supersonic turbulence \citep[e.g.,][]{kim2005,federrath2009}.

In the upper panel, the power spectra for the solenoidal and compressive components of the velocity, $P_v^{\rm sol}(k)$ and $P_v^{\rm comp}(k)$, are separately presented with solid and dotted lines. In contrast to $P_{\rho}(k)$, the differences in $P_v(k)$ with different forcings are not large. In particular, on small scales of $k/k_{\rm inj}\gtrsim$ a few, $P_v^{\rm comp}(k)$ as well as $P_v^{\rm sol}(k)$ are almost identical for the three cases of different forcings. This should be because the magnetic tension quickly converts compressive motions to Alfv\'en modes, and hence the solenoidal component of the velocity is efficiently generated even if the forcing is compressive. With the spatial extension of shock surfaces typically much smaller than $L_0$ (see Figure \ref{fig:fig2}), the frequency of shocks should be reflected mostly to $P_v^{\rm comp}(k)$ in the range of $k/k_{\rm inj}\gtrsim$ a few. With similar $P_v^{\rm comp}(k)$, below we see that the number of shock zones is similar in the models with different forcings. On the other hand, in $k/k_{\rm inj}\lesssim$ a few, $P_v^{\rm comp}(k)$ is larger for the compressive (green dotted) forcing model than for the solenoidal (red dotted) and mixed (blue dotted) forcing models; this is consistent with the large-scale distributions of shocks shown in Figure \ref{fig:fig2}. 

Another point to note is that while $P_v^{\rm comp}(k)$ has the slope close to $-2$, which is the slope of the Burgers spectrum in shock-dominated flows { \citep[e.g.,][]{kim2005,federrath2013}}, $P_v^{\rm sol}(k)$ is a bit flatter than the Kolmogorov spectrum. According to the Goldreich-Sridhar scaling, the Kolmogorov slope of $-5/3$ is expected for $P_v^{\rm sol}(k)$ in MHD turbulence, if the Alfv\'enic mode is dominant \citep{goldreich1995}. However, some simulations of ``incompressible'' MHD turbulence exhibited a slope close to $-3/2$ \citep[e.g.,][]{muller2005}, although the slope obtained in numerical simulations is controversial \citep[e.g.,][]{beresnyak2011}. Our simulations of ``compressible'' MHD turbulence produce $P_v^{\rm sol}(k)$ with a slope close to $\sim-1.2$ in the inertial range, indicating that the compressiblility is possibly involved. 

\begin{figure*}
\centerline{\includegraphics[width=1\textwidth]{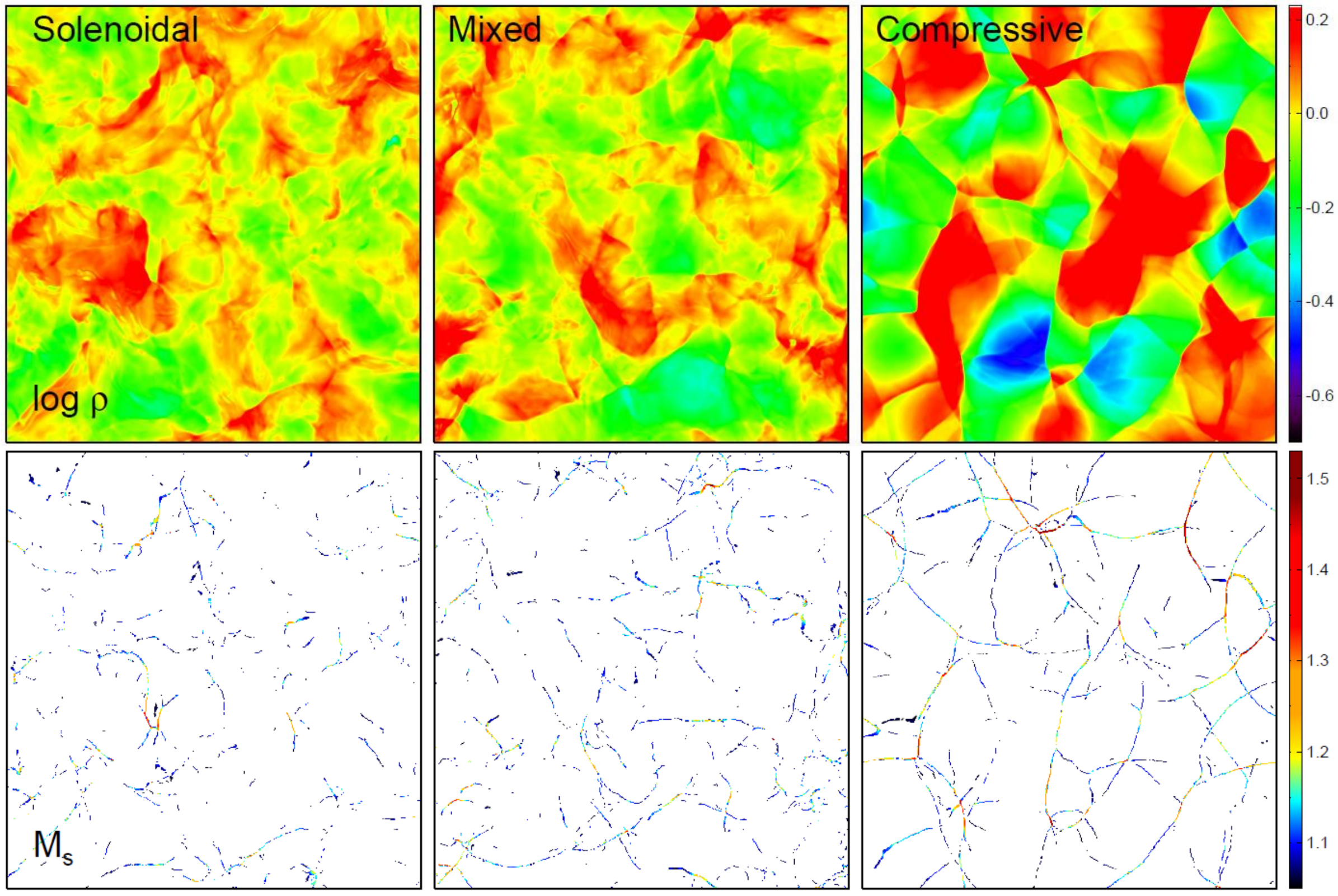}}
\vskip -0.1cm
\caption{2D slice images of $\log \rho/\rho_0$ (upper panels) and shock distribution color-coded by $M_s$ (lower panels) for the ICM turbulence models. Three high-resolution models, ICM-Sol-N512 (left), ICM-Mix-N512 (middle), and ICM-Comp-N512 (right), at $t=t_{\rm end}$ are shown. See Table \ref{tab:t1} for the model parameters.}
\label{fig:fig6}
\end{figure*}

\begin{figure}
\hskip -0.1cm
\centerline{\includegraphics[width=0.5\textwidth]{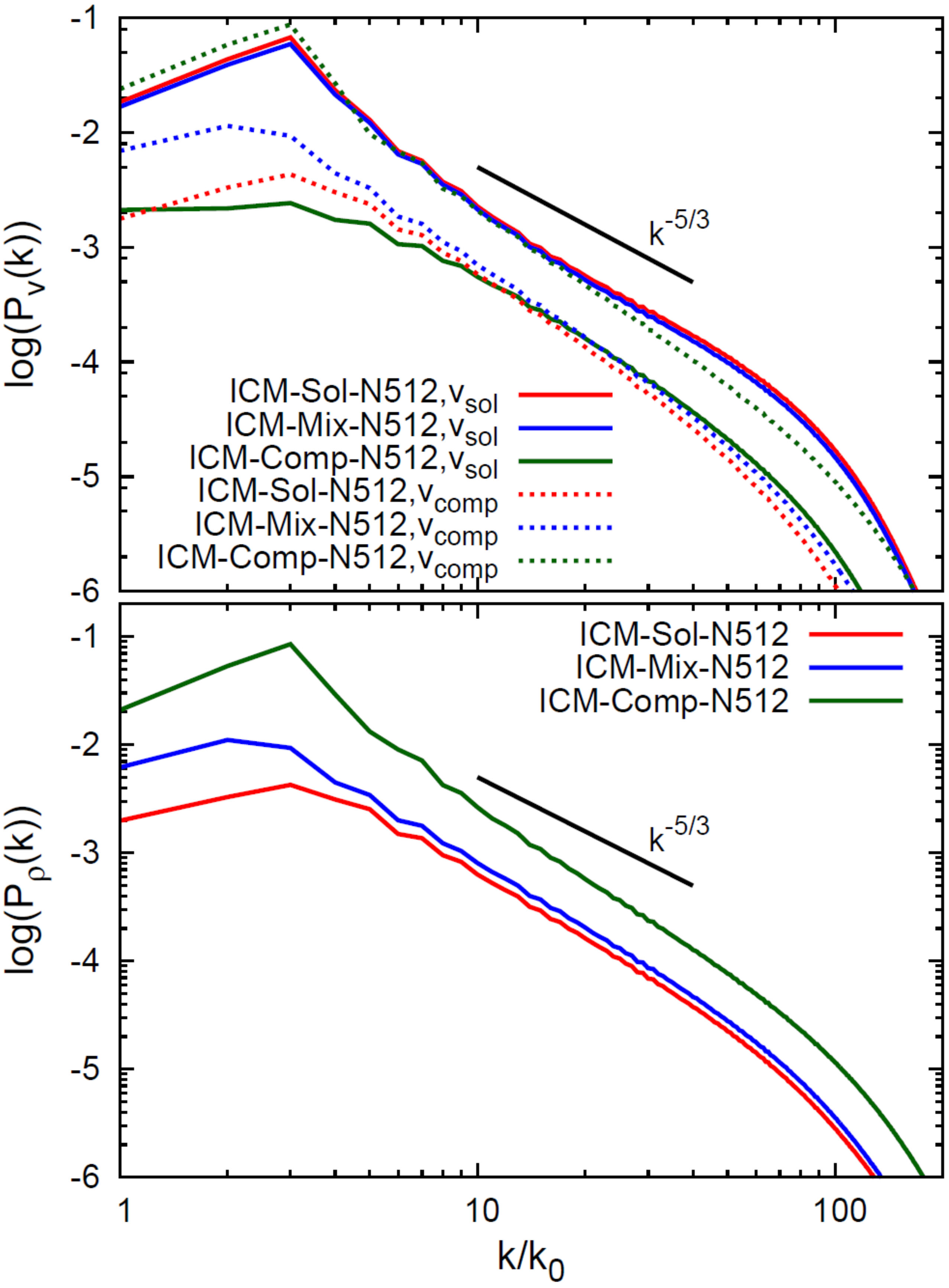}}
\vskip -0.2cm
\caption{{\it Upper panel:} Time-averaged power spectra of the solenoidal component of $\mbox{\boldmath$v$}$, $P_v^{\rm sol}(k)$ (solid lines), and the compressive component of $\mbox{\boldmath$v$}$, $P_v^{\rm comp}(k)$ (dotted lines), for the ICM turbulence models. {\it Lower panel:} Time-averaged power spectra of the density, $P_{\rho}(k)$, for the ICM turbulence models. The power spectra are calculated using the quantities at saturation ($15t_{\rm cross}\leq t\leq t_{\rm end}$). Three high-resolution models are shown. The black lines draw the Kolmogorov spectrum for comparison.}
\label{fig:fig7}
\end{figure}

The statistics of shocks are presented in Figure \ref{fig:fig4}. The upper panels plot the time-averaged PDFs of the sonic Mach number for fast and slow shocks, $(dN_{\rm fa}/dM_{s})/N_{\rm fa}$ and $(dN_{\rm sl}/dM_{s})/N_{\rm sl}$, averaged over the saturated period, for the six ISM models in Table \ref{tab:t1}. In all the cases of different forcings, the PDFs look similar; they peak at $M_{\rm peak}<M_{\rm turb}$ and decrease more or less exponentially at higher Mach numbers. The Mach numbers for fast shocks are much higher than those for slow shocks, as expected. In the N512} models the PDFs for fast shocks have $M_{\rm peak}\approx3.5\sim4.5$, and the characteristic Mach number $M_{\rm char}\sim 6$, if they are fitted to $\exp[-(M_s-1)/(M_{\rm char}-1)]$. By contrast, the PDFs for slow shocks have $M_{\rm peak}\approx1.1$ close to the lowest Mach number we identify, and the characteristic Mach number $M_{\rm char}\sim1.7$, although the distributions deviate from the exponential form at $M_s\gtrsim6$. The majority, $\sim80-85\%$, of fast shocks have the sonic Mach number less than the turbulent Mach number, while virtually all slow shocks have $M_s<M_{\rm turb}$, indicating that strong shocks with $M_s>M_{\rm turb}$ are relatively rare.

The most interesting point in the PDFs is that the difference due to different forcings is not significant, in contrast to the ICM turbulence (see the next subsection). As noted above, with the total magnetic energy is comparable to the kinetic energy,\footnote{The total magnetic energy is the sum of $E_B$ with $\delta\mbox{\boldmath$B$}$ in Figure \ref{fig:fig1} and the energy of the background magnetic field, \mbox{\boldmath$B_0$}, which is 10 in computational units.} it should be the consequence of strong magnetic tension; the incompressible, solenoidal mode dominantly appears in the flow velocity, in particular, for $k/k_{\rm inj}\gtrsim$ a few, regardless of forcings. Hence, the compressive mode, which is responsible for the formation of shocks, is subdominant and about the same for the three different forcing models.

Accordingly, the energy dissipation at shocks also shows only a weak dependence on forcing. The lower panels of Figure \ref{fig:fig4} plot the dissipation rates of the turbulent energy at fast and slow shocks as a function of the sonic Mach number, normalized to the energy injection rate, $(d\epsilon_{\rm fa}/dM_{s})/\epsilon_{\rm inj}$ and $(d\epsilon_{\rm sl}/dM_{s})/\epsilon_{\rm inj}$, which are averaged over the saturated period. Similarly to the Mach number PDF, the energy dissipation is dominated by shocks with small Mach numbers, while the contribution by high Mach number shocks decreases more or less exponentially. In the N512 models, $(d\epsilon_{\rm fa}/dM_{s})/\epsilon_{\rm inj}$ for fast shocks has peaks at $M_{\rm peak}\approx5.5\sim7$, and the characteristic Mach number $M_{\rm char}\sim 7.5$, if it is fitted to $\exp[-(M_s-1)/(M_{\rm char}-1)]$; $(d\epsilon_{\rm sl}/dM_{s})/\epsilon_{\rm inj}$ for slow shocks has peaks at $M_{\rm peak}\approx1.5$, and the characteristic Mach number $M_{\rm char}\sim 2$.

The total numbers of fast and slow shock zones, $N_{\rm fa}$ and $N_{\rm sl}$, normalized to the grid resolution, $n_g^2$, are listed in the eighth and ninth columns of Table \ref{tab:t1}, while the total energy dissipation rates at fast and slow shocks, integrated over the Mach number, $\epsilon_{\rm fa}$ and $\epsilon_{\rm sl}$, are listed in the tenth and eleventh columns. $N_{\rm fa}$ is several time larger than $N_{\rm sl}$ in our results. This is partly because only the shock zones with $M_s\geq1.05$ are counted; then, while most of fast shocks should be included, a substantial fraction of slow shocks might be missed since slow shocks could have even $M_s<1$. If shocks with $M_s<1.05$ are included, $N_{\rm sl}$ would be larger \citep[see][and PR2019]{lehmann2016}. On the other hand, the contribution of slow shocks with $M_s<1.05$ to the energy dissipation should not be significant. As a matter of fact, $\epsilon_{\rm fa}$ is much larger than $\epsilon_{\rm sl}$, as also noted in PR2019. 

The spatial frequency of shocks can be expressed in terms of the mean distance between shock surfaces,
$\langle d_{\rm shock}\rangle = L_0/(N_{\rm shock}/n_g^2) \propto N_{\rm shock}^{-1}$, where $N_{\rm shock}=N_{\rm fa}+N_{\rm sl}$. For fast and slow shocks of $M_s\geq1.05$ altogether, the mean distance is estimated to be $\langle d_{\rm shock}\rangle \sim0.3L_{\rm inj}$ with $L_{\rm inj}=L_0/2$ in the N512 models. The ratio of the energy dissipated at shocks and the injected energy, $\epsilon_{\rm shock}/\epsilon_{\rm inj}$, where $\epsilon_{\rm shock}=\epsilon_{\rm fa}+\epsilon_{\rm sl}$, is listed in the twelfth column of Table \ref{tab:t1}. It is $\sim15\%$ for the three different forcing models. This is roughly the fraction of the turbulent energy dissipated at shocks, while the rest, $\sim85\%$, of the turbulent energy should dissipate through the turbulent cascade.

Although the dependence on forcing is not large, there are still some differences in the shock statistics with different forcings. For instance, both $N_{\rm shock}$ and $\epsilon_{\rm shock}$ are larger in the solenoidal forcing model than in the compressive forcing model. This is partly because larger $\epsilon_{\rm inj}$ should be adopted for the solenoidal forcing model to achieve the same $M_{\rm turb}\approx10$ (see Table \ref{tab:t1}). { We point that smaller $\epsilon_{\rm inj}$ with compressive forcing would be an unexpected result, since compressive motions are expected to dissipate faster and hence compressive forcing would require larger $\epsilon_{\rm inj}$. Indeed, we see larger $\epsilon_{\rm inj}$ for the compressive case in the ICM turbulence with weak magnetic fields (Table \ref{tab:t1}). Again, this should be due to strong magnetic fields in the ISM turbulence. The magnetic tension seems to efficiently convert compressive motions to incompressive, Alfv\'en modes, and hence $\epsilon_{\rm inj}$ is not necessarily larger with compressive forcing.}

Also the shock statistics depend somewhat on the numerical resolution; $N_{\rm shock}$ is larger in the N512 models, while $\epsilon_{\rm shock}$ is larger in the N256 models. However, considering uncertainties in the identification of shock zones and the calculations of $M_s$ and $Q$, we regard that the statistics are reasonably resolution-converged. Comparing the shock statistics of the ISM-Sol-N512 model to those of the similar model in PR2019, 1024M7-b0.1 ($M_{\rm turb}\approx7$ and $\beta_0=0.1$), ISM-Sol-N512 has larger $N_{\rm shock}/n_g^2$ and $\epsilon_{\rm shock}/\epsilon_{\rm inj}$. This should be partly owing to the higher $M_{\rm turb}$ of ISM-Sol-N512, and also because the current WENO code with a higher order of accuracy seems to capture shocks better than the TVD code used in PR2019, particularly in complex flows with strong magnetic fields.

An interesting consequence of shocks is the density enhancement, which could have implications on the evolution of molecular clouds including the star formation rate (SFR). { As shown in Figure \ref{fig:fig2}, the density fluctuations are larger in the compressive forcing model; hence with more frequent occurrence of high density regions, the SFR would be enhanced. As a matter of fact, \citet{federrath2012} showed that the SFR would be about 10 times larger with compressive forcing than with solenoidal forcing in the turbulent ISM. We here examine how the forcing affects the density fluctuations at shocks as well as in the entire computational volume with the density PDF.}

In isothermal turbulence, the density PDF is often approximated as the lognormal distribution \citep[e.g.,][]{vazquez1994,padoan1997,passot1998}. { The standard deviation of the density distribution,} $\sigma$, is expected to be larger with compressive forcing than with solenoidal forcing. \citet{federrath2008}, for instance, showed that the radio of the compressive to solenoidal forcing cases is $\sigma_{\rm comp}/\sigma_{\rm sol}\sim3$, for the hydrodynamic turbulence with $M_{\rm turb}\sim5$. The value of $\sigma$ depends on $M_{\rm turb}$ and the magnetic field strength, or $\beta_0$ \citep[e.g.,][]{federrath2008,molina2012}, so does this ratio. The top panel of Figure \ref{fig:fig5} shows the PDFs of $\log\rho$ in the entire computational volume, averaged over the saturated period, for the N512 ISM models with three different forcing modes.\footnote{The common logarithm with base 10 is used to be consistent with other plots, while the natural logarithm was often used in previous studies.} In our simulations, { $\sigma_{\rm comp}/\sigma_{\rm sol}\approx2.1$,}; $\sigma_{\rm mix}$ for mixed forcing is similar to $\sigma_{\rm sol}$. With a larger $M_{\rm turb}\approx10$, the smaller ratio should be caused by strong magnetic fields. 

The middle and bottom panels of Figure \ref{fig:fig5} show the PDFs of $\log\rho$ at the preshock (dotted lines) and postshock (solid lines) regions, for fast and slow shocks, respectively. Again, compressive forcing leads to broader density distributions with larger $\sigma$'s in both the preshock and postshock regions than solenoidal forcing. In all the cases, $\sigma_{\rm comp}/\sigma_{\rm sol}$ is in the of range { $\approx2.2-2.9$} at both the preshock and postshock regions, and again $\sigma_{\rm mix}$ is similar to $\sigma_{\rm sol}$. These values around shocks are comparable to or slightly larger than $\sigma_{\rm comp}/\sigma_{\rm sol}$ for the entire volume.

\subsection{ICM Turbulence}\label{sec:s3.3}

Figure \ref{fig:fig6} shows the 2D slice maps of the density (upper panels) and the color-coded $M_s$ (lower panels) in the N512 ICM models with different forcings at $t=t_{\rm end}$. The density images exhibit expected distributions with different forcings, for instance, a higher intermittency in the compressive forcing model (right panel) than in the solenoidal forcing model (left panel). With $\beta_{\rm sat}\gg1$, i.e., subdominant magnetic fields at saturation, the turbulence should be almost hydrodynamic; hence the density distributions are similar to those of previous hydrodynamic studies, such as, \citet{federrath2008,federrath2009}, although the details would be different if $M_{\rm turb}$ is different. Our ICM turbulence model is subsonic with $M_{\rm turb}\approx0.5$, yet shocks are ubiquitous \citep[see also][]{porter2015}. Similarly to the density distributions, the shock distributions exhibit differences with different forcings; the distribution looks more organized with stronger shocks in the compressive forcing model than in the solenoidal forcing model. In the mixed forcing model (middle panel), the density and shock distributions show a bit larger intermittency and more shocks, compared to the solenoidal forcing model.

Figure \ref{fig:fig7} shows the power spectra for the solenoidal and compressive components of the flow velocity, $P_v^{\rm sol}(k)$ and $P_v^{\rm comp}(k)$ (upper panel), and the density, $P_{\rho}(k)$ (lower panel), averaged over the saturated period, for the three N512 ICM models. As in the ISM case, $P_{\rho}(k)$ is several to an order of magnitude larger in the compressive forcing model than in the other forcing models; $P_{\rho}(k)$ is a bit larger in the mixed forcing model than in the solenoidal forcing model. Even though there are discontinuities in the density distribution formed by shocks, in the ICM turbulence models with small $M_{\rm turb}$, $P_{\rho}(k)$ is steeper than the Kolmogorov spectrum in all the models with different forcings. This is consistent with the previous finding of \citet{kim2005} that in hydrodynamic turbulence, while $P_{\rho}(k)$ flattens as $M_{\rm turb}$ increases, the slope is less than $-5/3$ for $M_{\rm turb}\lesssim1$.

The upper panel shows that $P_v$ behaves differently from that of the ISM turbulence. While $P_v^{\rm sol}(k)$ dominates over $P_v^{\rm comp}(k)$ in all the wavenumbers in the solenoidal and mixed forcing models, $P_v^{\rm comp}(k)$ is much larger than $P_v^{\rm sol}(k)$ in the compressive forcing model, as was previously shown, for instance, in { \citet{federrath2011,porter2015}.} With negligible magnetic tension in the high $\beta$ ICM, the memory of forcing is persistent in the flows of fully developed turbulence. $P_v^{\rm comp}(k)$ is several times larger in the compressive forcing model than in the other forcing models, and $P_{\rm comp}(k)$ is a bit larger in the mixed forcing model than in the solenoidal forcing model. With larger $P_v^{\rm comp}(k)$, shocks would be more abundant in the compressive forcing model, as noted with the spatial distribution of shocks in Figure \ref{fig:fig6}. For all the models with different forcings, $P_v^{\rm comp}(k)$ is steeper than the Kolmogorov spectrum and has the slope close to $-2$, which is the slope of the shock-dominated Burgers spectrum. By contrast, $P_v^{\rm sol}(k)$ shows a slight concavity around $k\sim10-30$ in the solenoidal and mixed forcing models. It was argued that in the ICM turbulence, the power spectrum for the kinetic energy also has the concavity, but being compensated by the magnetic power, the power spectrum for the total energy roughly follows the Kolmogorov spectrum \citep[see, e.g.,][]{porter2015}. Hence, while the magnetic field is subdominant, it would still affect the flow motions and $P_v(k)$, especially in those solenoidal and mixed forcing models.

Unlike in the ISM turbulence models, virtually all the shocks formed in the ICM turbulence models are fast shocks, and the fraction of slow shocks is very small; for instance, $\sim1\%$ in the ICM-Sol-512 model and even smaller in the compressive and mixed forcing models. This is because the magnetic field strength decreases across slow shocks, and hence they can appear only when the  preshock magnetic field is sufficiently strong to satisfy the condition, $v_{{A}\parallel1}^2/c_s^2 > M_s^2$. With the small fraction and almost no contribution to the energy dissipation, below we do not further consider slow shocks, and present only the statistics of fast shocks, for the ICM models.

\begin{figure}
\hskip -0.1cm
\centerline{\includegraphics[width=0.5\textwidth]{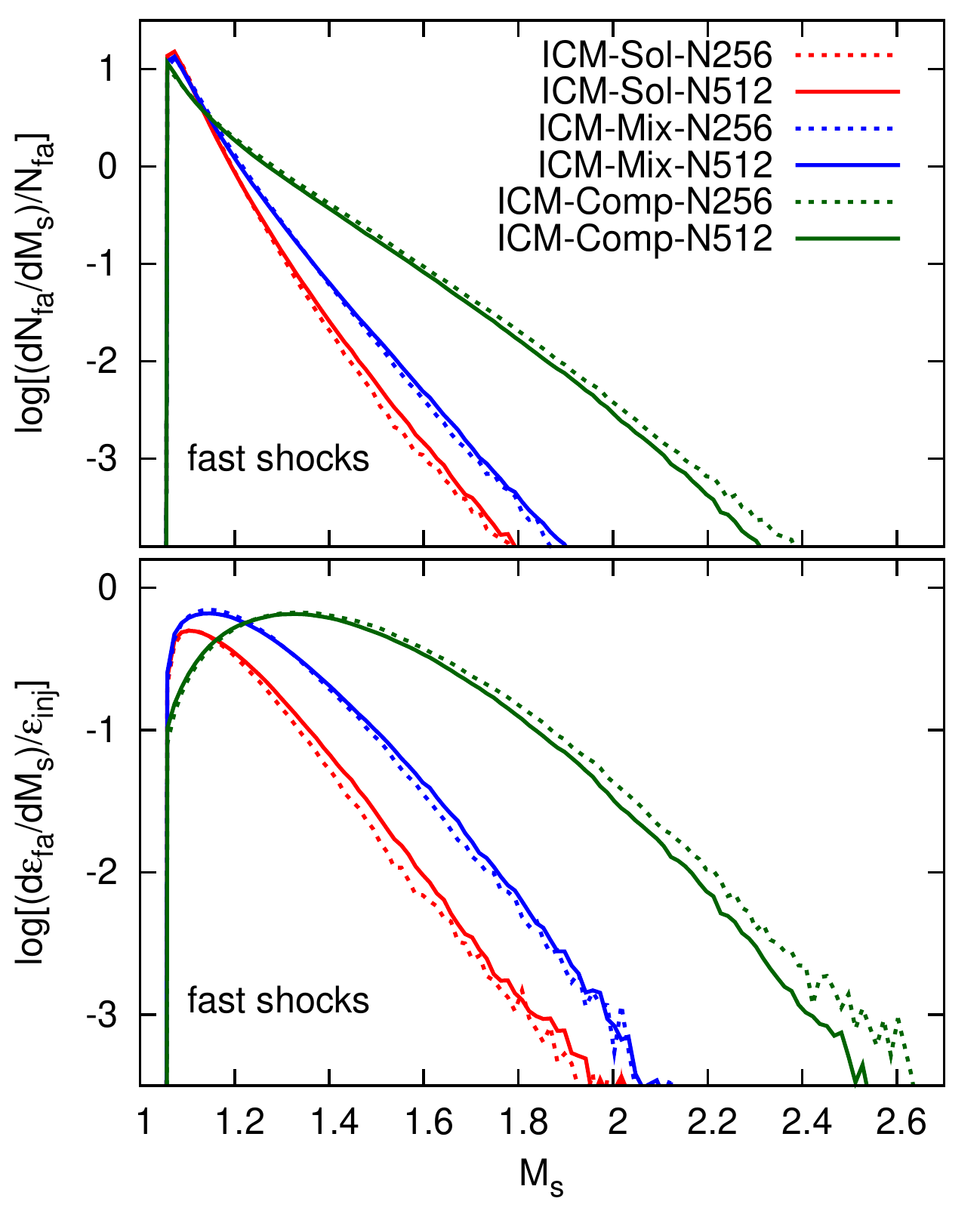}}
\vskip -0.2cm
\caption{Time-averaged PDFs of $M_s$ (upper panels) and time-averaged energy dissipation rates at shocks as a function of $M_s$ (lower panels) for fast shocks in the ICM turbulence models. The distributions are calculated using the quantities at saturation ($15t_{\rm cross}\leq t\leq t_{\rm end}$). All the six ICM models in Table \ref{tab:t1} are shown.}
\label{fig:fig8}
\end{figure}

Figure \ref{fig:fig8} presents the statistics of fast shocks. The upper panel shows the PDFs of the sonic Mach number, $(dN_{\rm fa}/dM_{s})/N_{\rm fa}$, averaged over the saturated period, for the six ICM models in Table \ref{tab:t1}. As in the ISM cases, the PDFs peak at small Mach numbers and decrease more or less exponentially at high Mach numbers. However, unlike in the ISM cases, the PDFs differ substantially with different forcings; there are more shocks with higher Mach numbers in the compressive forcing model (green lines). With $M_{\rm turb}<1$, the peak occurs at $M_{\rm peak}\approx1.05$, the lowest Mach number we identify, in all the forcing models. But when the PDFs are fitted to $\exp[-(M_s-1)/(M_{\rm char}-1)]$, the characteristic Mach numbers are $M_{\rm char}\sim 1.07$, $1.08$, and $1.13$ for the solenoidal, mixed, and compressive forcing models, respectively. This reveals that with compressive forcing, shocks with higher compression and higher $M_s$ are more abundant. We note that these characteristic Mach numbers agree well with those of \citet{porter2015}, who quoted $M_{\rm char}\sim 1.08$ and $1.125$ for the solenoidal and compressive forcing cases, although different numerical codes and different shock identification schemes are used.

The lower panel of Figure \ref{fig:fig8} plots the dissipation rate of the turbulent energy at fast shocks normalized to the energy injection rate, $(d\epsilon_{\rm fa}/dM_{s})/\epsilon_{\rm inj}$, which are averaged over the saturated period. Again the distributions peak at small Mach numbers and decrease more or less exponentially at high Mach numbers in all the cases of different forcings, whereas it shifts to higher Mach numbers in the compressive forcing model. The peak and characteristic Mach numbers are $M_{\rm peak}\approx1.10$, $1.14$, and $1.33$, and $M_{\rm char}\sim 1.11$, $1.12$, and $1.16$, for the solenoidal, mixed, and compressive forcing models, respectively.\footnote{Although $(dN_{\rm fa}/dM_{s})/N_{\rm fa}$ and $(d\epsilon_{\rm fa}/dM_{s})/\epsilon_{\rm inj}$ look quite different with different forcings in Figure \ref{fig:fig8}, the values of $M_{\rm peak}$ and $M_{\rm char}$ are similar. This is because they are estimated as a function of $M_s-1$; the values of $M_{\rm peak}-1$ and $M_{\rm char}-1$ are sufficiently different with different forcings.}

The total number of fast shock zones, $N_{\rm fa}$, normalized to the grid resolution, $n_g^2$, and the total energy dissipation rate at fast shocks, integrated over the Mach number, $\epsilon_{\rm fa}$, are given in Table \ref{tab:t1}. As expected from the PDFs in Figure \ref{fig:fig8}, $N_{\rm fa}$ is substantially larger in the compressive forcing model than in the other forcing models, and also $N_{\rm fa}$ is noticeably larger in the mixed forcing model than in the solenoidal model. Specifically, $N_{\rm fa}$ of ICM-Comp-N512 and ICM-Mix-N512 is $\sim1.8$ and $\sim1.3$ times $N_{\rm fa}$ of ICM-Sol-N512. The mean distance between shock surfaces is $\langle d_{\rm shock}\rangle\sim0.31$, $0.24$, and $0.17L_{\rm inj}$ in the N512 ICM models with solenoidal, mixed, and compressive forcings, respectively. An interesting point is that fast shocks are more frequent in our ICM turbulence models than in the ISM models, although $M_{\rm turb}$ is 20 times smaller. While strong magnetic fields in the ISM models limit the gas compression and hence inhibit the formation of shocks, subdominant magnetic fields in the ICM models do not significantly affect the occurrence of shocks. 

Accordingly, $\epsilon_{\rm fa}$ is much larger in the compressive forcing model than in the other forcing models. For instance, $\epsilon_{\rm fa}$ of ICM-Comp-N512 is $\sim7.5$ times $\epsilon_{\rm fa}$ of ICM-Sol-N512. On the other hand, the energy injection rate, $\epsilon_{\rm inj}$, differs substantially with different forcings, and $\epsilon_{\rm inj}$ of ICM-Comp-N512 is $\sim2.4$ times $\epsilon_{\rm inj}$ of ICM-Sol-N512. As a result, the normalized energy dissipation rate, $\epsilon_{\rm fa}/\epsilon_{\rm inj}$, is only $\sim3.1$ times larger in ICM-Comp-N512 than in ICM-Sol-N512. And $\epsilon_{\rm fa}/\epsilon_{\rm inj}$ is $\sim1.8$ times larger in ICM-Mix-N512 than in ICM-Sol-N512. The fraction of the turbulent energy dissipated at shocks is estimated to be $\epsilon_{\rm shock}/\epsilon_{\rm inj}\sim11\%$, $\sim19\%$, and $\sim33\%$ in the N512 ICM models with solenoidal, mixed, and compressive forcings, respectively. Compared to the fractions in the ISM turbulence, the fraction in the solenoidal forcing model is slightly smaller, but the fractions in the compressive and mixed forcing models are larger.
 
The dependence of the shock statistics on numerical resolution is rather weak in the ICM turbulence models, as can be seen in Figure \ref{fig:fig8} and Table \ref{tab:t1}. For instance, $N_{\rm fa}/n_g^2$ and $\epsilon_{\rm shock}/\epsilon_{\rm inj}$ are $\sim6\%$ larger in ICM-Sol-N512 than in ICM-Sol-N256, and the differences are even smaller in the other forcing models. Again, considering uncertainties in the estimation of those quantities, the statistics would be regarded to be reasonably resolution-converged. 

The shock statistics of the ICM-Sol-N512 model may be compared to those of 1024M0.5-b10 in PR2019 ($M_{\rm turb}\approx0.5$ and $\beta_0=10$). The shock frequency, $N_{\rm fa}/n_g^2$, is larger in ICM-Sol-N512 than in 1024M0.5-b10, partly because shocks with $M_s\geq1.05$ are counted in this work, while those with $M_s\geq1.06$ are included in PR2019. Also in ICM-Sol-N512 with a much weaker initial magnetic field ($\beta_0=10^6$), $\beta_{\rm sat}$ is about 10 times larger than in 1024M0.5-b10, that is, the magnetic energy at saturation is about 10 times smaller, and hence more shocks form. However, the shock dissipation fraction, $\epsilon_{\rm shock}/\epsilon_{\rm inj}$, is actually smaller in ICM-Sol-N512. This is because the term involving $v_{\rm A}$ in Equation (\ref{eq-Q}), which represents the energy dissipation through the magnetic field, makes a relatively small contribution to $\epsilon_{\rm shock}$ in ICM-Sol-N512, while it is substantial in 1024M0.5-b10, especially at weak shocks.

\begin{figure}
\hskip -0.1cm
\centerline{\includegraphics[width=0.5\textwidth]{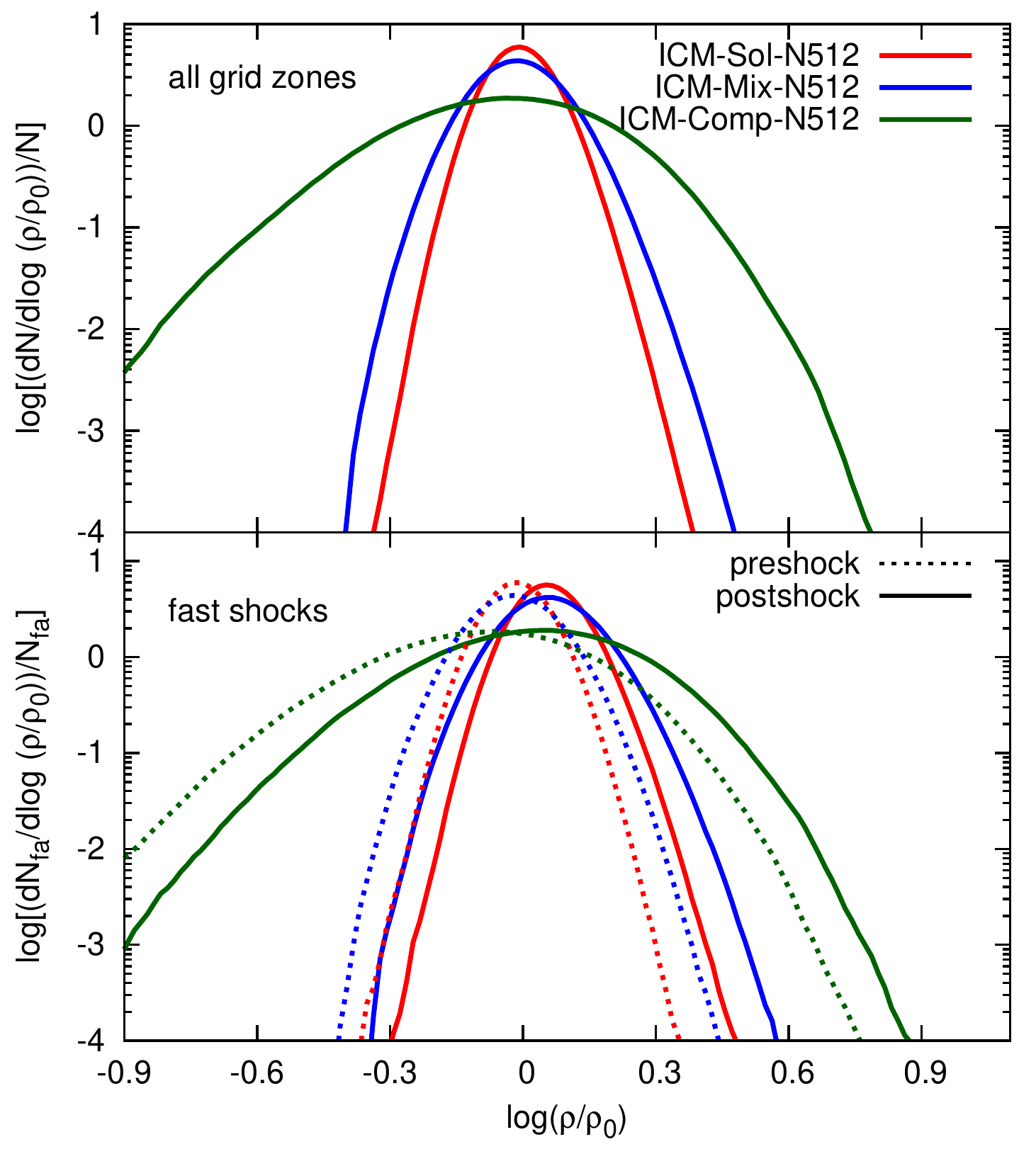}}
\vskip -0.2cm
\caption{Time-averaged PDFs of the gas density in all the computational volume (upper panel) and the preshock and postshock gas densities for fast shocks (lower panel) in the ICM turbulence models. The PDFs are calculated using the quantities at saturation ($15t_{\rm cross}\leq t\leq t_{\rm end}$). Three high-resolution ICM models are shown.}
\label{fig:fig9}
\end{figure}

The upper panel of Figure \ref{fig:fig9} displays the PDFs of $\log\rho$ in the entire computational volume, averaged over the saturated period, for the three N512 ICM models with different forcings. As in the ISM cases, compressive forcing leads to a larger density contrast with larger $\sigma$ than solenoidal forcing. In our simulations, { $\sigma_{\rm comp}/\sigma_{\rm sol}\approx3.1$}; this value is roughly the same as that of \citet{federrath2008} for the hydrodynamic turbulence with $M_{\rm turb}\sim5$, indicating $\sigma_{\rm comp}/\sigma_{\rm sol}$ would not be very sensitive to $M_{\rm turb}$, as long as the magnetic field is subdominant. The width of the PDF in the mixed forcing model is somewhat larger than in the solenoidal forcing model, { with $\sigma_{\rm mix}/\sigma_{\rm sol}\approx1.4$.} The bottom panel plots the PDFs of $\log\rho$ at the preshock and postshock regions of fast shocks. While the peaks of the PDFs are located at low and high densities for the preshock and postshock regions, respectively, the widths look similar to those for the whole computational volume. Quantitatively, { $\sigma_{\rm comp}/\sigma_{\rm sol}\approx3.0$ and $2.9$} for the preshock and postshock regions, respectively, and $\sigma_{\rm mix}/\sigma_{\rm sol}\approx1.3$ for both the preshock and postshock regions.

\section{Summary and Discussion}\label{sec:s4}

We have simulated supersonic turbulence in the low-$\beta$ ISM and subsonic turbulence in the high-$\beta$ ICM, using a high-order accurate MHD code based on the finite-difference WENO scheme \citep{jiang1996,jiang1999}. In particular, we have employed different forcings to { drive} turbulence, considering solenoidal and compressive modes as well as a mixture of those modes with the 2:1 ratio. The amplitude of forcings is adjusted so that, in the saturated state, the turbulent Mach number reaches $M_{\rm turb}\approx 10$ for the ISM turbulence with the initial plasma beta $\beta_0=0.1$, and $M_{\rm turb}\approx 0.5$ for the ICM turbulence  with $\beta_0=10^6$. We have then analyzed the simulation data, focusing on the statistics of shocks, that is, the PDF of the shock Mach number, $M_s$ and the dissipation rates of the turbulent energy at shocks, $\epsilon_{\rm fa}$ and $\epsilon_{\rm sl}$, for fast and slow shocks, respectively. We have also examined the power spectra of the solenoidal and compressive components of the flow velocity and the density, and evaluated the PDF of the density in the preshock and postshock regions. The main results are summarized as follows.

1. In the ISM turbulence models, the shock statistics overall show only weak dependence on forcings. The PDFs of $M_s$ look similar, regardless of forcings; they peak at the Mach numbers less than $M_{\rm turb}$, and decrease more or less exponentially at higher Mach numbers. The majority ($\sim85\%$) of shocks have $M_s<M_{\rm turb}$. Shocks are slightly more frequent in the solenoidal forcing model than in the compressive forcing model, partly because a higher energy injection rate, $\epsilon_{\rm inj}$, is required for the solenoidal forcing model.
The shock frequency in the mixed forcing model is almost the same as that of the solenoidal forcing model. The mean distance between the surfaces of shocks with $M_s\geq1.05$ is estimated to be $\langle d_{\rm shock}\rangle \sim0.3L_{\rm inj}$, regardless of forcings. The dissipation rate of the turbulent energy at shocks, $\epsilon_{\rm shock}$, is also slightly larger in the solenoidal forcing model than in the compressive forcing model. However, the fraction of the turbulent energy dissipated at shocks, $\epsilon_{\rm shock}/\epsilon_{\rm inj}$, is the other way around, that is, $\epsilon_{\rm shock}/\epsilon_{\rm inj}$ is slightly larger in the compressive forcing model. Yet, in all the cases, $\epsilon_{\rm shock}/\epsilon_{\rm inj}$ is estimated to be $\sim15\%$. The rest of the turbulent energy should be dissipated through turbulent cascade.

2. On the contrary, in the ICM turbulence models, the shock statistics exhibit a strong dependence on forcing. The PDFs of $M_s$ have peaks at $M_{\rm peak}\sim 1$ in all the models, but they have broader widths in the compressive forcing model than in the other forcing models; hence, shocks are more frequent and also stronger on average in the compressive forcing model. This is partly because the compressive driving produces shocks more efficiently and also because $\epsilon_{\rm inj}$ is larger in the compressive forcing model. The mean distance between the surfaces of shocks with $M_s\geq1.05$ is $\langle d_{\rm shock}\rangle \sim0.31$, $0.24$, and $0.17L_{\rm inj}$ for the solenoidal, mixed, and compressive forcing models, respectively. The shock dissipation rate, $\epsilon_{\rm shock}$, is substantially larger in the compressive forcing model; it is $\sim7.5$ times $\epsilon_{\rm shock}$ of the solenoidal forcing model. However, $\epsilon_{\rm inj}$ is also larger in the compressive forcing model, and hence the ratio $\epsilon_{\rm shock}/\epsilon_{\rm inj}$ differs less. The fraction of the turbulent energy dissipated at shocks, $\epsilon_{\rm shock}/\epsilon_{\rm inj}$, is estimated to be $\sim11\%$, $\sim19\%$, and $\sim33\%$ for the solenoidal, mixed, and compressive forcing models, respectively.

3. In the ISM turbulence models, both fast and slow shocks are present. While slow shocks could be as frequent as fast shocks \citep[see][and PR2019]{lehmann2016}, they account only for $\sim20\%$ of the shocks identified with $M_s>1,05$. Slow shocks are weaker and also dissipate less energy than fast shocks. Hence, the energy dissipation at slow shocks is estimated to be $\sim2-3\%$ of that at fast shocks. In the ICM turbulence models, almost all of the identified shocks are fast shocks. The fraction of slow shocks is only $\sim1\%$ in the solenoidal forcing model, and even smaller in the other forcing models. Accordingly, the energy dissipation at slow shocks is negligible.

\begin{figure*}
\centerline{\includegraphics[width=0.75\textwidth]{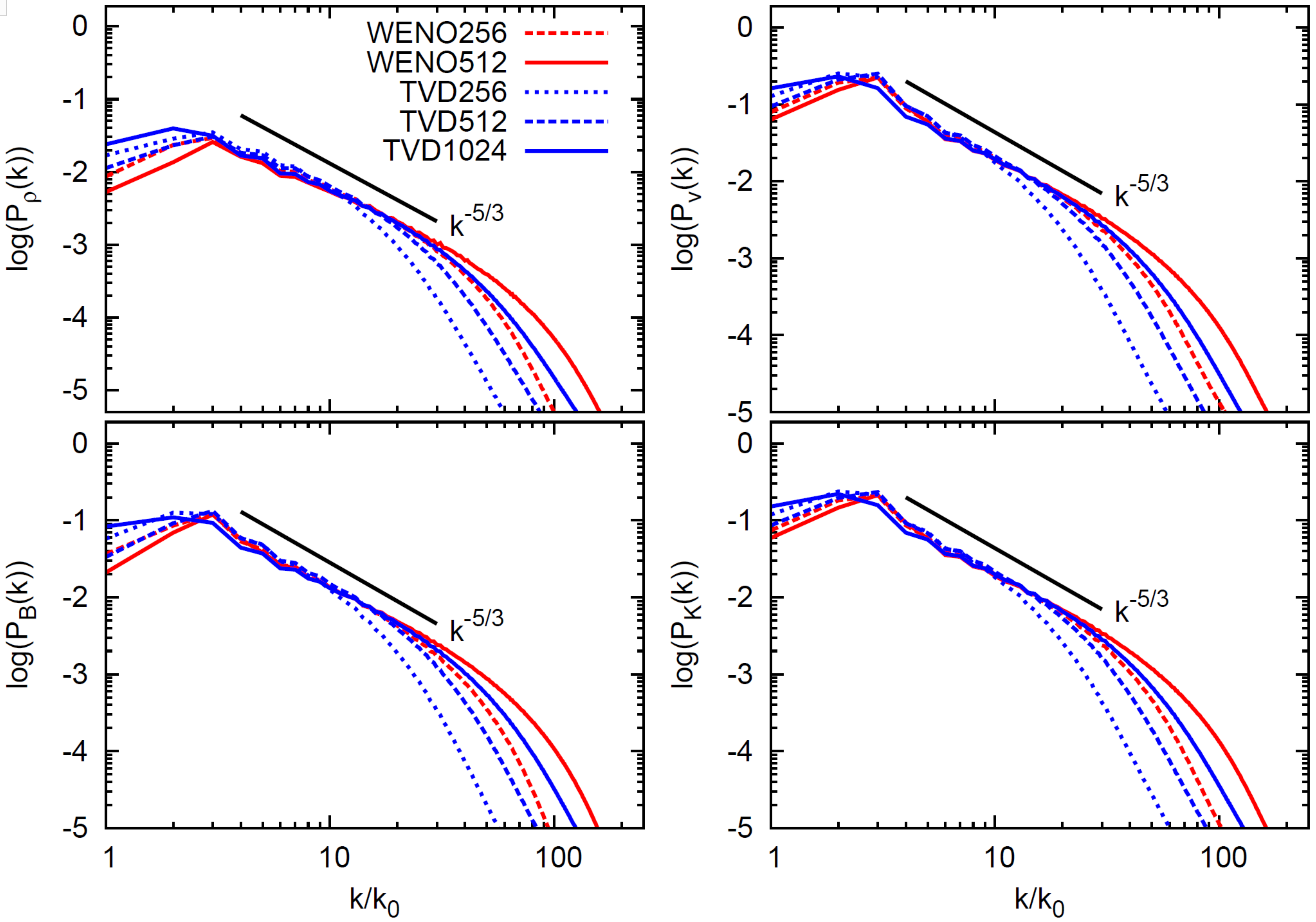}}
\vskip -0.1cm
\caption{Time-averaged power spectra of the density, $P_{\rho}(k)$ (top-left), velocity, $P_{v}(k)$ (top-right), magnetic field, $P_B$ (bottom-left), and kinetic energy, $P_K$ (bottom-right), for the MHD turbulence with $M_{\rm turb}\approx1$ and $\beta_0=1$ from simulations using the WENO code with $256^3$ and $512^3$ grid zones (red lines) and the TVD code with $256^3$, $512^3$,and $1024^3$ grid zones (blue lines). The spectra at saturation are shown.
\label{fig:fig10}}
\end{figure*}

4. The density PDF is often fitted to the lognormal distribution { \citep[e.g.,][]{vazquez1994,passot1998,federrath2008}}, and the standard deviation of the density distribution, $\sigma$, depends on forcing. In the ISM turbulence models, the ratio of the compressive to solenoidal forcing cases is estimated to be { $\sigma_{\rm comp}/\sigma_{\rm sol}\approx2.2-2.9$} at the preshock and postshock regions. This is comparable to or slightly larger than { $\sigma_{\rm comp}/\sigma_{\rm sol}\approx2.1$} estimated for the whole computational volume. By contrast, in the ICM turbulence models, { $\sigma_{\rm comp}/\sigma_{\rm sol}\approx3.0$ and 2.9} at the preshock and postshock regions, which is about the same as the ratio for the whole computational volume, { $\sigma_{\rm comp}/\sigma_{\rm sol}\approx3.1$}. 

5. The power spectra of the density, $P_{\rho}(k)$, and the flow velocity, $P_v(k)$, exhibit the behaviors that reflect the shock statistics. In the ISM turbulence models, $P_{\rho}$ is larger in the compressive forcing model than in the other forcing models, revealing more intermittent density distribution. Contrastingly, $P_v$ shows only a weak dependence on forcing in small scales of $k/k_{\rm inj}\gtrsim$ a few, which is consistent with the weak dependence of the shock statistics on forcing. In the ICM turbulence models, both $P_v$ and $P_{\rho}$ depend sensitively on forcings. In particular, $P_{\rho}$ as well as $P_v^{\rm comp}$ are larger in the compressive forcing model, manifesting the more intermittent density distribution and the larger population of shocks.

As shock is one of the important aspects of turbulence, the quantification of shock frequency and energy dissipation could help us understand better the physical processes in the ISM and ICM, as well as observations of, such as spectral lines in the ISM and radio synchrotron in the ICM. We leave investigations of those to future works.

\begin{acknowledgments}
This work was supported by the National Research Foundation (NRF) of Korea through grants 2016R1A5A1013277, 2020R1A2C2102800, and 2020R1F1A1048189. Some of simulations were performed using the high performance computing resources of the UNIST Supercomputing Center.
\end{acknowledgments}

\appendix
\section{An isothermal MHD code based on the WENO scheme}

Simulations have been carried out using a code based on a weighted essentially non-oscillatory (WENO) scheme. WENO is one of upwind schemes, which is designed to achieve a high-order accuracy in smooth regions and keep the essentially non-oscillatory property near discontinuities; hence, it should accurately reproduce the nonlinear dynamics in turbulent flows. The basic idea of the WENO scheme lies in the adaptive reconstruction of physical fluxes \citep[see][for a review]{shu2009}. \citet{jiang1996} formulated a 5th-order accurate finite difference (FD) WENO scheme for hydrodynamics, in which the fluxes estimated at the cell center are used to produce the reconstructed fluxes at the cell interfaces with weight functions. The MHD extension was described in \citet{jiang1999}. Our code for isothermal MHDs in Equations (\ref{eq-rho})-(\ref{eq-ene}) employs this 5th-order accurate FD WENO scheme. For the time integration, the classical, 4th-order accurate Runge-Kutta (RK) method is employed \citep[e.g.,][]{shu1988,jiang1996}. The $\mbox{\boldmath$\nabla$}\cdot\mbox{\boldmath$B$}$ condition is enforced using a constrained transport (CT) scheme, described in \citet{ryu1998}. Viscous and resistive dissipations are not explicitly modeled.

For comparison and also test purposes, we have performed simulations for the MHD turbulence of $M_{\rm turb}\approx1$ and $\beta_0=1$ using the WENO code, as well as the 2nd-order accurate TVD code \citep{ryu1995,ryu1998}, which was employed in PR2019. The turbulence has been derived with solenoidal forcing. 

Figure \ref{fig:fig10} shows the power spectra of the density, $P_{\rho}(k)$, the flow velocity, $P_v(k)$, the kinetic energy, $P_K(k)$, and the magnetic energy, $P_B(k)$, averaged over the saturated period. The power spectra for different codes and different resolutions exhibit similar amplitudes and slopes in the inertial range, demonstrating the statistical agreement of turbulent flows in different simulations. All the power spectra follow more or less the Kolmogorov slope in the inertial range. An important point is that with the same resolution, the WENO code has the inertial range wider than the TVD code, implying the higher order nature of the WENO code. As a matter of fact, the Kolmogorov slope stretches over the wider range in the WENO512 case than in the TVD1024 case. Although there could be the issue of bottleneck effect \citep[see, e.g.,][]{falkovich1994} in high-resolution simulations, it seems to indicates that the WENO512 simulation would reproduce small-scale, turbulent flow structures as well as, or even better than, the TVD1024 simulation.\footnote{We point out that for the same resolution, the computational cost of the WENO code is $\sim10-15$ times of that of the TVD code, partly offsetting the advantage of the high-order accurate WENO code.}

\bibliography{turb}{}
\bibliographystyle{aasjournal}

\end{document}